\documentclass[aps,prl,twocolumn,showkeys,superscriptaddress,nofootinbib,floatfix]{revtex4-2}
\usepackage{amsmath,amssymb,bm,graphicx,booktabs,microtype,xcolor}
\usepackage{needspace,placeins}
\usepackage{etoolbox}
\AtBeginEnvironment{thebibliography}{\interlinepenalty=10000}
\usepackage[hidelinks]{hyperref}
\makeatletter
\let\frontmatter@footnote@produce\frontmatter@footnote@produce@footnote
\makeatother
\graphicspath{{figures/}}
\newcommand{\ket}[1]{|#1\rangle}
\newcommand{\bra}[1]{\langle#1|}
\newcommand{\braket}[2]{\langle#1|#2\rangle}

\begin{document}

\title{Chiral Floquet Selection of an Algebraic Rydberg Bound State in the Continuum}
\author{X.~Z.~Zhang}
\email{zhangxz@tjnu.edu.cn}
\affiliation{College of Physics and Materials Science, Tianjin Normal University, Tianjin 300387, China}
\affiliation{Interdisciplinary Center, Tianjin Normal University, Tianjin 300387, China}
\author{X.-F.~Zhang}
\email{xfzhang@sust.edu.cn}
\affiliation{School of Physics \& Information Science, Shaanxi University of Science and Technology, Xi'an 710021, China}
\author{W.-M.~Liu}
\email{wmliu@aphy.iphy.ac.cn}
\affiliation{College of Physics and Materials Science, Tianjin Normal University, Tianjin 300387, China}
\affiliation{Beijing National Laboratory for Condensed Matter Physics, Institute of Physics, Chinese Academy of Sciences, Beijing 100190, China}

\begin{abstract}
Bound states in the continuum (BICs) remain localized despite the presence of extended modes at the same energy.
In driven systems with long-range couplings, the absence of radiation does not by itself determine how a bound state decays in space.
We study a Hermitian Rydberg spin waveguide coupled to a four-site plaquette, where a winding drive phase combines rotation with time translation to forbid propagating guide channels in one symmetry sector.
The resulting eigenstate remains embedded in the continuum of other sectors and carries an inverse-cube waveguide amplitude, with coherent contributions from direct plaquette--guide coupling and long-range guide propagation.
Pair-state calculations determine the microscopic couplings in the axial geometry, while coherent Floquet mixing sets the weight of the dominant circulating plaquette mode.
The bound state thus retains the spatial signature of the long-range coupling even when radiation is forbidden.
\end{abstract}

\maketitle

\emph{Introduction.}---A bound state in the continuum (BIC) is a normalizable mode embedded in a spectrum of extended states \cite{vonNeumann1929,Fano1961,Stillinger1975,Friedrich1985,Miroshnichenko2010,Hsu2016,Sadreev2021}. Such states occur in wave and quantum systems in which coupling to a continuum does not necessarily lead to radiation \cite{Fan2002,Marinica2008,Plotnik2011,Yang2014,Koshelev2019}. Algebraically localized embedded states have also been demonstrated in engineered waveguide lattices \cite{Corrielli2013}, showing that embedding need not imply exponential localization. When the coupling to the continuum is long ranged, one must determine the spatial decay of the bound state as well as the condition for vanishing radiation.

Periodic driving makes the available radiation channels depend on both energy and harmonic index. Floquet BICs have been identified in periodically driven lattices \cite{Longhi2013}. Drive control of localization also appears in flux-induced wave-packet confinement in non-Hermitian rings \cite{ZhangSong2019} and in non-Hermitian Floquet edge modes \cite{ZhangGong2020}. Quasienergy folding creates continuum replicas, and the harmonic label can carry symmetry information in the enlarged Floquet space of Fourier components (Sambe space) \cite{Shirley1965,Sambe1973,Grifoni1998,Bukov2015,Goldman2014,Eckardt2015,Eckardt2017,Lignier2007,Jotzu2014,Oka2019,Rudner2020}. A state can therefore share its quasienergy with extended modes while the symmetry sector labeled by its angular and Fourier indices contains no propagating guide channel. Rydberg arrays provide a microscopic setting for controlled spin Hamiltonians \cite{Saffman2010,Weimer2010,Saffman2016,Browaeys2020,Morgado2021}. Individually controlled arrays realize prescribed geometries and correlated states \cite{Labuhn2016,Barredo2016,Endres2016,Bernien2017,Omran2019}; resonant dipolar exchange supplies the long-range, angle-dependent coupling \cite{deLeseleuc2017,Ravets2015}; and periodic driving enables Hamiltonian engineering in isolated Rydberg spin systems \cite{Geier2021}. Long-range interactions reshape the propagation of correlations and excitations \cite{Hauke2013,Jurcevic2014,Richerme2014,FossFeig2015,Gong2023}. In the system studied here, the long-range guide shapes the bound state's spatial profile even when symmetry forbids radiation.

\begin{figure}[t]
\includegraphics[width=\columnwidth]{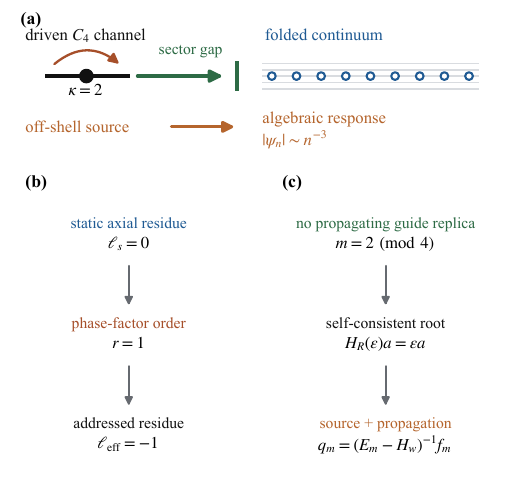}
\caption{Floquet symmetry separates radiation from the long-range bound response.
(a) A winding-driven \(C_4\) mode belongs to a joint sector whose accessible guide replicas are gapped even though other sectors form a folded continuum; the nonpropagating coupling remains long ranged.
(b) The axial static channel has \(\ell_s=0\), and phase-factor order \(r=1\) couples to the angular component \(\ell_{\rm eff}=-1\).
(c) The complete state, including its waveguide component, is nonradiating because no propagating guide replica is available in its \(\kappa=2\) sector, while its bound guide component obeys \(q_m=(E_m-H_w)^{-1}f_m\).
An embedded eigenstate must also satisfy the real self-consistent eigenvalue equation.}
\label{fig:mechanism}
\end{figure}

We consider a closed Hermitian system consisting of a winding-driven Rydberg plaquette coupled to an axial dipolar spin waveguide. The axial geometry supplies a long-range coupling with only the static angular component \(\ell_s=0\), while the winding factor changes the angular label seen by the continuum. Radiation is forbidden when the state's symmetry sector contains no propagating guide channel, but a bound state must also satisfy the real self-consistent eigenvalue equation. Its nonpropagating guide response combines the direct plaquette--guide source with propagation through the long-range guide. The result is an algebraic Floquet BIC whose nonradiating waveguide amplitude follows \(|\psi_n|\sim n^{-3}\). We determine the microscopic couplings using the Alkali Rydberg Calculator (ARC) \cite{Sibalic2017,Robertson2021} at the actual separations and angles. Floquet symmetry therefore excludes propagating replicas while coupling to the same dipolar continuum gives the bound state its spatial tail (Fig.~\ref{fig:mechanism}).

\emph{Model and Floquet selection.}---We work in the single-excitation sector of a four-site Rydberg plaquette coupled to a one-dimensional spin waveguide. The plaquette sites are labeled by angular coordinates \(\theta_j=\pi j/2\), and the waveguide atoms are placed on an axis through the plaquette center. The four angular modes of the plaquette are labeled by \(\ell=0,\pm1,2\) modulo four. In the absence of driving the two circulating modes \(\ell=\pm1\) are degenerate, while the axial waveguide geometry selects a static plaquette-to-waveguide coupling vector with only \(\ell_s=0\) angular content. The drive is \(H(t)=F(t)H_0F^\dagger(t)\), where \(H_0\) is the static matrix and \(F(t)\) multiplies plaquette site \(j\) by \(e^{i\phi_j(t)}\), leaving the guide unchanged. Thus both internal plaquette links and plaquette--guide contacts acquire winding phases. The phase contains the two harmonics \(z_1\sin(\Omega t-\theta_j+\varphi_1)\) and \((z_2/2)\sin(2\Omega t-2\theta_j+\varphi_2)\), with parameters specified in the Supplemental Material, Sec.~I. A Fourier component of the phase factor of order \(r\), proportional to \(e^{ir\Omega t-ir\theta_j}\), carries angular index \(-r\). If \(J_{\ell_s n}^{(0)}\) denotes the static angular projection of the plaquette-to-site-\(n\) coupling, the continuum channel selected by this factor is
\begin{equation}
\ell_{\rm eff}=\ell_s-r\pmod 4 .
\label{eq:sambe_product}
\end{equation}
For the axial geometry, the static channel is \(\ell_s=0\), while \(r=1\) couples to \(\ell_{\rm eff}=-1\), as summarized in Fig.~\ref{fig:mechanism}(b). The phase-factor order \(r\) is not the Floquet-state harmonic. With \(u(t)=\sum_m u_m e^{-im\Omega t}\), the joint rotation--time-translation label is \(\kappa=m-\ell\pmod4\). Its sector-resolved extended spectrum is
\begin{equation*}
\sigma_{\rm ess}\!\left(H_{\rm F}^{(\kappa)}\right)=\bigcup_{m=\kappa\;({\rm mod}\;4)}\left(\mathcal B-m\Omega\right),
\end{equation*}
where \(\mathcal B\) is the undriven guide band. For the parameters considered here, an algebraically localized Floquet BIC occurs in the \(\kappa=2\) sector; this is the branch studied below. Guide states have \(\ell=0\), so this state has accessible guide harmonics only at \(m=2\pmod4\). Their physical sideband energies are \(E_m=\epsilon+m\Omega\); the nearest allowed energies \(E_{\pm2}\) lie outside \(\mathcal B\), whereas the in-band \(m=0,1\) replicas belong to other sectors. Its quasienergy can therefore lie in the global continuum even though no guide mode propagates in its own symmetry sector. The Fourier-sign convention, the \(C_4\) angular basis, and the static-channel projection are derived in the Supplemental Material, Sec.~III, and the Floquet-Feshbach conditions in Sec.~II.

Within this state's symmetry sector, the angular weights follow from hybridization among levels coupled by the drive and shifted by the guide. Different angular components can therefore coexist at different harmonics of the same nonradiating Floquet state.

For the single-tone \(L_w=321\) state, the dominant components in the representation where the drive acts on site energies (the onsite representation) are \((\ell,m)=(2,0),(-1,1),(0,2)\). Their detunings are comparable to the drive coupling, so the eigenstate mixes these components rather than occupying only the addressed harmonic. Using the reference exchange scale \(J_*\) defined in Supplemental Material, Sec.~I, their bare Floquet energies are \(0.5731J_*\), \(-0.2478J_*\), and \(-0.6443J_*\). The closed \(m=2\) guide block contributes a dispersive self-energy \(0.3530J_*\), and the fundamental drive couples neighboring harmonics with magnitude \(0.6655J_*\). The amplitudes \(b_{\ell,m}\) and \(a_{\ell,m}\) are the angular--harmonic components of \(u_{\rm on}=F^\dagger u\) and \(u\), respectively, and \(c_r\) is defined by \(e^{i\phi_j(t)}=\sum_r c_r e^{ir\Omega t-ir\theta_j}\). Transforming the hybridized state back to the physical phase representation gives
\[
a_{\ell,m}=\sum_r c_r b_{\ell+r,m+r}.
\]
For \(a_{-1,1}\), the three leading terms have magnitudes \(0.2640\), \(0.1560\), and \(0.3159\) with a common phase; including the remaining harmonics gives \(|a_{-1,1}|^2\simeq0.61817\). Coherent hybridization therefore enhances the \(\ell=-1\) weight, while the joint symmetry forbids radiation by excluding propagating guide channels (Supplemental Material, Sec.~III and Fig.~S1).

\emph{Algebraic embedded branch.}---The Floquet modes are obtained from the unitary evolution over one drive period,
\begin{equation}
U(T)\ket{u_\nu}=e^{-i\epsilon_\nu T}\ket{u_\nu}.
\label{eq:floquet_unitary}
\end{equation}
As the guide is extended, the same branch retains finite plaquette weight \(P_{\rm plaq}\), target-chirality weight \(P_{\chi_-}\), and inverse participation ratio (IPR), while the waveguide-like folded spectrum becomes denser.
The periodic quasienergy distance \(\Delta_{\rm ext}\) measures its proximity to that extended subset [Fig.~\ref{fig:branch}(a,b)].
The three local quantities are full-state snapshots at \(t=0\), defined in Supplemental Material, Sec.~I.
The corresponding infinite-guide state lies in a gap of its own symmetry sector and in the continuum of other sectors (Supplemental Material, Sec.~IV).

Eliminating the waveguide with a Feshbach projection separates the radiation condition from the bound spatial response. An analogous energy-dependent self-energy reduction has been used for superconducting circuits coupled to structured waveguides \cite{Zhou2026}. In Floquet-Sambe space, let \(P\) and \(Q=1-P\) project onto the plaquette and guide, and let \(H_{\rm F}\) be the Floquet operator. The elimination gives a self-energy with Hermitian dispersive part \(\Delta_F\); the columns of \(W_{\rm open}\) are the coupling vectors to independent propagating channels. With \(H_{F,PP}=PH_{\rm F}P\), the localized component \(\ket{a_B}=P\ket{u_B}\) of a BIC at quasienergy \(\epsilon_B\) obeys
\begin{equation}
\begin{aligned}
W_{\rm open}^{\dagger}(\epsilon_B)\ket{a_B}&=0,\\
\bigl[H_{F,PP}+\Delta_F(\epsilon_B)\bigr]\ket{a_B}
&=\epsilon_B\ket{a_B}.
\end{aligned}
\label{eq:bic_criterion}
\end{equation}
The first line excludes radiation, while the second determines the self-consistent bound eigenstate. In the present construction, joint symmetry supplies the first condition by excluding every accessible propagating guide replica. The corresponding guide component obeys
\begin{equation}
(\epsilon_B-QH_{\rm F}Q)Q\ket{u_B}=QH_{\rm F}P\ket{a_B}.
\label{eq:feshbach_tail}
\end{equation}
The radiation condition removes the projection of the source \(QH_{\rm F}P\ket{a_B}\) onto propagating outgoing modes, but it does not make the source vanish. Writing \(q_m\) and \(f_m\) for the guide response and contact source at harmonic \(m\), one has \(q_m=(E_m-H_w)^{-1}f_m\), with \(E_m=\epsilon_B+m\Omega\). For a uniform inverse-cube guide, the small-momentum dispersion contains a \(k^2\ln|k|\) nonanalyticity, giving an off-band full-line Green function \(G_n^{\rm line}(E)\propto n^{-3}\). The response of the physical semi-infinite guide combines this long-range propagation with the direct plaquette--guide source, so both contributions enter the leading spatial coefficient. The derivation and its scope are given in the Supplemental Material, Sec.~V.

In the physical semi-infinite guide, the contact source and guide propagation enter the asymptotic response through two complex amplitudes. Here \(\omega_w=(H_w)_{nn}\), \(t_d=(H_w)_{n,n+d}\) is the real guide hopping, which depends only on the site separation, and \(J_\infty=\lim_{d\to\infty}d^3t_d\). For the nonoscillatory, absolutely summable response specified in Supplemental Material, Sec.~V, if \(f_{n,m}=A_m n^{-3}+o(n^{-3})\) and \(q_{n,m}=B_m n^{-3}+o(n^{-3})\), the off-band guide equation gives
\[
\begin{aligned}
\left(E_m-\omega_w-2\sum_{d\ge1}t_d\right)B_m
&=A_m+J_\infty\mathcal M_m,\\
\mathcal M_m&=\sum_{n\ge1}q_{n,m}.
\end{aligned}
\]
The first term on the right is the direct contact source and the second is long-range propagation from the occupied guide sites near the plaquette. For the converged single-tone \(L_w=641\) state, the finite-guide ratios \(J_\infty\mathcal M_m^{(L_w)}/A_m\), with \(\mathcal M_m^{(L_w)}=\sum_{n=1}^{L_w}q_{n,m}\), are \(+0.0896\) for \(m=2\) and \(-0.0484\) for \(m=-2\), showing that propagation can either reinforce or oppose the direct source. Exact cancellation of the two terms would remove the leading inverse-cube coefficient. The inverse-cube envelope thus follows from the combined source and guide response (Supplemental Material, Sec.~V).

The computed Floquet profiles approach an inverse-cube envelope, consistent with the geometrical reference,
\begin{equation}
\begin{aligned}
K_{\rm geo}(n)&\equiv(R_{n_0}/R_n)^3\sim n^{-3},\\
|\braket{s_n}{u_B(0)}|&\sim n^{-3}\qquad (n\gg1).
\end{aligned}
\label{eq:tail_law}
\end{equation}
Here \(R_n\) is the plaquette-to-guide separation and \(n_0\) is the normalization site, so \(K_{\rm geo}\) is a distance-only reference; the Floquet source also depends on the harmonic amplitudes of the state, as derived in Supplemental Material, Sec.~V. The waveguide amplitude has the same leading power, so the corresponding probability is summable as \(n^{-6}\). The exponent therefore reflects the combined dipolar source and guide response after radiation has been excluded by symmetry.

Figure~\ref{fig:branch} shows the spectrum and spatial profile of this branch.
Its plaquette weight, target chirality, and IPR remain essentially unchanged over \(L_w=21\)--641.
At \(L_w=641\), the nearest extended-state spacings are \(\Delta_{\rm ext}/J_*\simeq1.09\times10^{-4}\) and \(6.41\times10^{-4}\) for the single-tone and biharmonic drives, respectively; the relation to the infinite-guide spectrum is derived in Supplemental Material, Sec.~IV.
The amplitude exponent remains \(\alpha=3.006\)--\(3.010\) across three interior windows at this size (Supplemental Material, Sec.~VIII).
The target chirality follows from coherent hybridization rather than from Eq.~(\ref{eq:sambe_product}) alone: for the single-tone state, \(P_{\chi_-}(0)\simeq0.615\) and \(\overline P_{\chi_-}\simeq0.618\), with their small difference arising from interference between harmonics during a drive period (Supplemental Material, Sec.~III).

To connect the finite branch to an infinite-guide eigenstate, we bound its residual under the infinite Floquet operator. Let \(u_{L,M}\) be the normalized Floquet trial state in the onsite representation, set to zero outside the retained sites and harmonics, at \(L=L_w=641\) and harmonic cutoff \(M=16\), with finite-system quasienergy \(\epsilon_{L,M}\). Self-adjointness gives
\[
{\rm dist}\!\left(\epsilon_{L,M},\sigma(H_{\rm F}^{(\kappa)})\right)
\le \left\|(H_{\rm F}^{(\kappa)}-\epsilon_{L,M})u_{L,M}\right\|.
\]
For the two waveforms, the full infinite-operator residuals are bounded by \(5.24\times10^{-7}J_*\) and \(5.39\times10^{-7}J_*\), respectively, while the distance to the \(\kappa=2\) continuum exceeds \(0.818J_*\). Each residual interval lies within this sector's gap and inside the continuum of another sector. It therefore contains at least one eigenvalue of the effective Floquet operator whose normalizable eigenstate is embedded in the continuum of another sector (Supplemental Material, Sec.~IV). The chirality and spatial decay follow from the Floquet mode in Fig.~\ref{fig:branch} and the source-response analysis in Sec.~V of the Supplemental Material.

We follow the localized branch through five values of \(z_2\) between 0 and 0.1, keeping all static and remaining drive parameters fixed.
The adjacent-state squared overlap remains above \(0.99996\) (Supplemental Material, Sec.~III), consistent with a continuous branch in this symmetry-preserving waveform family.

\begin{figure}[t]
\includegraphics[width=\columnwidth]{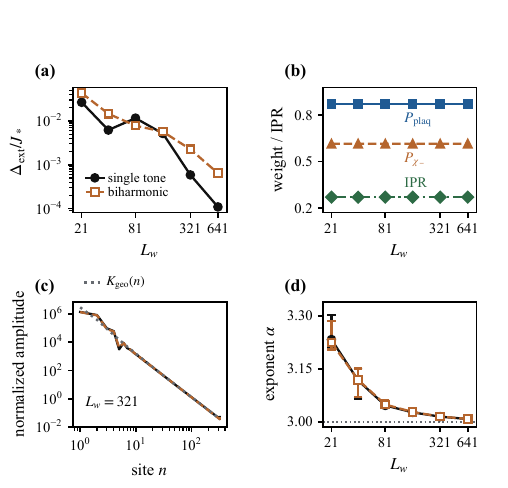}
\caption{Spectral proximity and algebraic localization of the same Floquet branch.
(a) Periodic quasienergy distance \(\Delta_{\rm ext}\) to the waveguide-like extended subset for the single-tone and biharmonic drives.
(b) Plaquette weight, target-chirality weight, and IPR of the single-tone branch.
(c) \(L_w=321\) snapshot amplitudes for both waveforms and the geometrical reference \(K_{\rm geo}\), each normalized at \(n=108\); colors and markers follow panel (a).
(d) Amplitude exponents fitted in interior window \(\mathcal W_B\), with whiskers spanning windows \(\mathcal W_A\)--\(\mathcal W_C\) of Supplemental Material, Sec.~VIII.
The guide has open ends and uses \(\Omega/J_*=1.21\), \(z_1=1.1\), and \(z_2=0\) or \(0.1\).}
\label{fig:branch}
\end{figure}

This is a Floquet form of symmetry protection: the conserved label combines rotation and time translation, so the state need not occupy a single angular mode at a fixed time. Unlike Friedrich--Wintgen cancellation between open resonant paths \cite{Friedrich1985}, the present joint sector has no propagating guide replica; the self-consistent real equation determines the embedded eigenstate within that sector.

\begin{figure}[!t]
\includegraphics[width=\columnwidth]{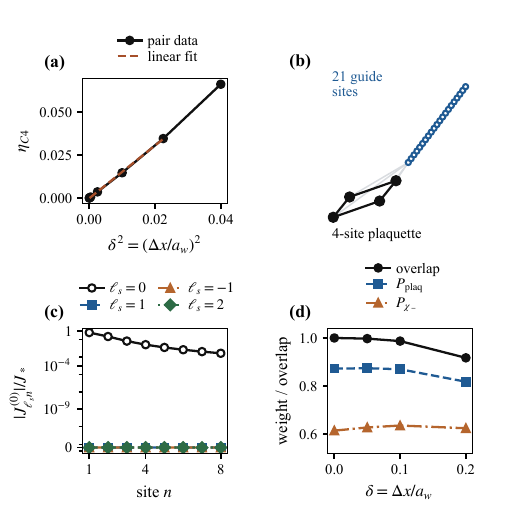}
\caption{Axial angular channel and its symmetry-breaking response.
(a) A transverse displacement produces a quadratic onset of the unwanted-to-target static angular-channel power ratio \(\eta_{C_4}\); the dashed line fits the small-displacement regime.
(b) Schematic view of the 25-atom geometry, with the four-site plaquette enlarged for visibility.
(c) At the axial point, the direct ARC exchange has a finite \(\ell_s=0\) component while the \(\ell_s=\pm1,2\) components vanish to numerical precision.
(d) The finite Floquet branch remains continuously connected to the axial state over the plotted displacement range. Panels (a,c) use the static contacts to the first eight guide sites at \(B=20\) G; panel (b) shows the 25-atom geometry with a 21-site guide. The Floquet-state calculation in panel (d) uses \(L_w=161\), the same field, and the single-tone drive (Supplemental Material, Sec.~IX).}
\label{fig:geometry}
\end{figure}

\emph{Microscopic geometry and symmetry.}---The axial Rydberg geometry in Fig.~\ref{fig:geometry} realizes the static angular channel used above.
The selected 25-atom device has plaquette side \(s\simeq12.4128~\mu{\rm m}\), axial offset \(h\simeq3.9489~\mu{\rm m}\), and guide spacing \(a_w\simeq12.7781~\mu{\rm m}\).
All 300 pair exchanges are evaluated at their actual separations and angles with ARC \cite{Sibalic2017,Robertson2021}, at \(B=20\) G.
For the longer guides, the pair exchange is evaluated directly at each distinct separation and angle.
The fixed-energy reduction, prescribed onsite terms, and full-precision geometry are specified in Supplemental Material, Secs.~I, VI, and VII.

At the axial point, the four plaquette--guide contacts are equivalent and their static angular projection contains only \(\ell_s=0\).
The \(r=1\) winding factor addresses \(\ell_{\rm eff}=-1\) through Eq.~(\ref{eq:sambe_product}), while the complete state retains the joint-sector radiation condition derived above.
The angle dependence of the signed pair exchange gives the effective spin model this angular structure. The contact source and propagation along the guide together determine the tail.

The axial selection is exact at the level of the four contact amplitudes \(C_{jn}=\bra{p_j}H_0\ket{s_n}\). If symmetry makes \(C_{jn}=c_n^{(0)}\) independent of the plaquette site \(j\), their angular projection is
\[
J_{\ell_s n}^{(0)}
=\frac{1}{2}\sum_{j=0}^{3}e^{-i\ell_s\pi j/2}C_{jn}
=2c_n^{(0)}\delta_{\ell_s,0}^{(4)}.
\]
Here \(\delta_{\ell_s,0}^{(4)}\) is one for \(\ell_s=0\) modulo four and zero otherwise. Thus the forbidden angular components vanish because of the geometry itself, while the pair calculation fixes the signed magnitude of the surviving contact. A transverse shift \(\delta=\Delta x/a_w\) destroys this equivalence. Reflection symmetry groups the displaced contacts as \((a_n,b_n,b_n,a_n)\), where \(a_n,b_n\) denote static contact amplitudes, not Floquet-state coefficients. For a smoothly varying pair exchange with nonzero axial coupling, the ratio of unwanted to target angular-channel power reduces to
\[
\eta_{C_4}(\delta)
=\frac{\sum_{n=1}^{8}|a_n(\delta)-b_n(\delta)|^2}
{\sum_{n=1}^{8}|a_n(\delta)+b_n(\delta)|^2}
=c_\delta\delta^2+O(\delta^4).
\]
The sums cover the same eight guide sites as Fig.~\ref{fig:geometry}(a), and \(c_\delta\) is the quadratic coefficient at the axial point. Reversing \(\delta\) exchanges \(a_n\) and \(b_n\), so both powers are even and the next correction is fourth order. The quadratic onset in Fig.~\ref{fig:geometry}(a) therefore follows from the breaking of the angular selection rule (Supplemental Material, Secs.~VII and IX).

The static angular mixing is accompanied by a continuous evolution of the finite Floquet state from its axial reference over the plotted range, with squared overlap with the axial state \(0.918\) at \(\delta=0.20\) [Fig.~\ref{fig:geometry}(d)]. The complete displacement dependence is given in Supplemental Material, Sec.~IX; the infinite-guide BIC statement applies to the symmetry-preserving axial geometry.

\FloatBarrier
\emph{Discussion.}---An embedded state can remain nonradiating while retaining an algebraic response to the long-range continuum.
The joint Floquet symmetry excludes propagating guide channels from the state's sector, while the Hermitian eigenproblem determines the bound solution.
Direct contact coupling and propagation through the dipolar guide then produce its algebraic envelope.
Coherent Floquet hybridization determines its chirality, and the axial Rydberg geometry supplies the required static angular channel.

The waveform deformation preserves the combined symmetry and hence the allowed guide harmonics, whereas transverse displacement admits angular components excluded on the axis.
The distinction extends beyond the particular Rydberg parameters used here. In a driven long-range continuum, symmetry can forbid radiation at the state's quasienergy while the remaining coupling to the guide produces a spatial tail. Its coefficient depends on both the source and propagation through the guide. The existence of a self-consistent embedded eigenstate and its decay exponent must be determined for each system. The spatial decay of a Floquet BIC can therefore reveal how long-range coupling acts even when radiation is forbidden.

\emph{Acknowledgments.}---X.Z.Z. acknowledges support from the National Natural Science Foundation of China under Grants No. 12275193 and No. 12675024. X.-F.Z. acknowledges support from the National Natural Science Foundation of China under Grant No. 12475004 and the Youth Innovation Team of Shaanxi Universities. W.-M.L. acknowledges support from the National Key R\&D Program of China under Grants Nos. 2024YFF0726700, 2021YFA1400900, and 2021YFA0718300; the National Natural Science Foundation of China under Grants Nos. 12334012, 12234012, and 52327808; the Space Application System of China Manned Space Program; and the Elite Revitalizing Inner Mongolia Program (2025TGL05).

\emph{Data availability.}---The data and scripts that support the figures are available from the corresponding authors upon reasonable request.

\providecommand{\newblock}{}

\onecolumngrid
\clearpage
\setcounter{equation}{0}
\setcounter{figure}{0}
\setcounter{table}{0}
\setcounter{section}{0}
\setcounter{subsection}{0}
\setcounter{subsubsection}{0}
\renewcommand{\theHequation}{supp.\arabic{equation}}
\renewcommand{\theHfigure}{supp.\arabic{figure}}
\renewcommand{\theHtable}{supp.\arabic{table}}
\renewcommand{\theHsection}{supp.\arabic{section}}
\renewcommand{\theHsubsection}{\theHsection.\arabic{subsection}}
\renewcommand{\theHsubsubsection}{\theHsubsection.\arabic{subsubsection}}
\pdfbookmark[0]{Supplemental Material}{supplemental-material}
\begin{center}
{\large\bfseries Supplemental Material}\par\medskip
{\normalsize\itshape Chiral Floquet Selection of an Algebraic Rydberg Bound State in the Continuum}
\end{center}
\medskip

\renewcommand{\thefigure}{S\arabic{figure}}
\renewcommand{\theequation}{S\arabic{equation}}
\setcounter{secnumdepth}{1}

\noindent {\color{black}This Supplemental Material gives the model, derivations, and numerical details supporting the main text. Section~I defines the Hamiltonian, units, and observables. Sections~II and III derive the Floquet-Feshbach condition, the joint space-time symmetry, and the chirality of the bound state. Section~IV connects the finite-size branch to an embedded eigenstate of the infinite guide. Section~V derives the source and propagation contributions to the algebraic tail under the stated asymptotic assumptions. Sections~VI and VII describe the atomic reduction, its range of validity, and the axial geometry. Sections~VIII and IX discuss numerical convergence and the response to a transverse displacement.}

\section{Definitions, projectors, and spectral measures}
\label{sec:sm_definitions}

{\color{black}We set \(\hbar=1\) and use angular-frequency units for \(H\), the quasienergy \(\epsilon\), and the drive frequency \(\Omega\). ARC pair matrices instead store cyclic frequencies in GHz. With \(\nu_*=0.006500146305671334~{\rm GHz}\), define \(J_*=2\pi\nu_*\), with the GHz-to-Hz conversion understood when time is expressed in seconds. The dimensionless variables are
\begin{equation}
h_0=\frac{H_0}{J_*}
=\frac{H_0^{\rm cyc}}{\nu_*},\qquad
\tau=J_*t,\qquad \omega=\frac{\Omega}{J_*}=1.21,\qquad
T_\tau=\frac{2\pi}{\omega}.
\label{eq:sm_units}
\end{equation}
Here \(H_0^{\rm cyc}\) is the cyclic-frequency pair-exchange matrix after the prescribed onsite terms are set. Thus \(i\partial_\tau\psi=h(\tau)\psi\); there is no additional \(2\pi\) in this dimensionless propagator. The physical cyclic drive frequency is \(\nu_{\rm drive}=\omega\nu_*\simeq7.865~{\rm MHz}\). The quoted digits in \(\nu_*\) define the numerical energy scale used below, and quasienergies are reported as \(\epsilon/J_*\). All angular-momentum labels of the four-site plaquette are defined modulo four; \(\ell=-1\) and \(\ell=3\) denote the same representation.}

In the single-excitation Hilbert space, a basis vector is either a plaquette excitation,
\begin{equation}
\ket{p_j}=p_j^\dagger\ket{0},\qquad j=0,1,2,3,
\end{equation}
or a waveguide excitation,
\begin{equation}
\ket{s_n}=s_n^\dagger\ket{0},\qquad n=1,\ldots,L_w .
\end{equation}
The state can therefore be written as
\begin{equation}
\ket{\psi(t)}=
\sum_{j=0}^{3} a_j(t)\ket{p_j}
+
\sum_{n=1}^{L_w} \psi_n(t)\ket{s_n}.
\label{eq:sm_state_expansion}
\end{equation}
The plaquette and waveguide projectors are
\begin{equation}
P=\sum_{j=0}^{3}\ket{p_j}\bra{p_j},\qquad
Q=\sum_{n=1}^{L_w}\ket{s_n}\bra{s_n},\qquad
P+Q=1.
\label{eq:sm_projectors}
\end{equation}
{\color{black}The normalized Floquet mode is denoted by \(u(t)\), with physical state \(e^{-i\epsilon t}u(t)\). Unless an average is explicitly indicated, all populations refer to the phase representation at the common time origin \(t=0\):
\begin{equation}
P_{\rm plaq}=\bra{u(0)}P\ket{u(0)},\qquad
P_{\chi_-}=|\braket{\ell=-1}{u(0)}|^2,\qquad
{\rm IPR}=\sum_r|u_r(0)|^4.
\label{eq:sm_snapshot_observables}
\end{equation}
The normalized angular ket \(\ket{\ell}\) is defined in Sec.~III, and \(r\) runs over all plaquette and guide sites. These are full-state weights, not conditional probabilities divided by \(P_{\rm plaq}\). The extended-subset distance \(\Delta_{\rm ext}\), defined in Eq.~(\ref{eq:sm_extended_distance}), has angular-frequency units; plotted values use \(\Delta_{\rm ext}/J_*\). It measures finite-chain spectral proximity, not a linewidth.}

 {\color{black}The static single-excitation matrix has the block form
\begin{equation}
H_0=\begin{pmatrix}K&C\\ C^\dagger&H_w\end{pmatrix},\qquad
C_{jn}=\bra{p_j}H_0\ket{s_n}.
\label{eq:sm_static_blocks}
\end{equation}
The Hermitian \(4\times4\) matrix \(K\) contains all six plaquette links, including opposite corners. The guide matrix \(H_w\) contains every retained guide-guide link, and \(C\) contains all plaquette-guide contacts with their signed complex amplitudes. The imposed onsite values are \(K_{jj}/J_*=1.08-0.01170571942455195\) and \((H_w)_{nn}/J_*=0.70\). The onsite detunings are prescribed independently of the pair-extraction diagonal shifts.}

{\color{black}For \(\theta_j=\pi j/2\), define
\begin{align}
\phi_j(t)&=z_1\sin(\Omega t-\theta_j+\varphi_1)
 +\frac{z_2}{2}\sin(2\Omega t-2\theta_j+\varphi_2),
\label{eq:sm_drive_phase}\\
F(t)&=\sum_{j=0}^3 e^{i\phi_j(t)}\ket{p_j}\bra{p_j}+Q .
\label{eq:sm_phase_operator}
\end{align}
The two waveforms have \(z_1=1.1\), \(\varphi_1=0.49\pi\), and \(\varphi_2=0\); only \(z_2\) changes from zero to \(0.1\). The second-harmonic phase amplitude is therefore \(0.05\) at the biharmonic endpoint. The complete driven Hamiltonian is
\begin{equation}
H_{\rm ph}(t)=F(t)H_0F^\dagger(t)
=\begin{pmatrix}F_PK F_P^\dagger&F_PC\\ C^\dagger F_P^\dagger&H_w\end{pmatrix},
\qquad F_P=PFP.
\label{eq:sm_full_driven}
\end{equation}
In particular, \((H_{\rm ph})_{p_jp_l}=K_{jl}e^{i(\phi_j-\phi_l)}\) and \((H_{\rm ph})_{p_js_n}=C_{jn}e^{i\phi_j}\). Internal plaquette bonds are driven along with the contacts. The Hamiltonian is closed and Hermitian, with no terminal loss or absorber.}

{\color{black}The exact onsite representation follows by writing \(\psi_{\rm ph}=F\psi_{\rm on}\). Substitution in the Schr\"odinger equation gives
\begin{align}
i\dot F\,\psi_{\rm on}+iF\,\dot\psi_{\rm on}&=FH_0\psi_{\rm on},\\
i\dot\psi_{\rm on}
&=(H_0-iF^\dagger\dot F)\psi_{\rm on},\qquad
F^\dagger\dot F=i\sum_j\dot\phi_j\ket{p_j}\bra{p_j}.
\end{align}
Consequently,
\begin{align}
H_{\rm on}(t)&=H_0+\sum_j\dot\phi_j(t)\ket{p_j}\bra{p_j},\\
\dot\phi_j(t)&=\Omega z_1\cos(\Omega t-\theta_j+\varphi_1)
+\Omega z_2\cos(2\Omega t-2\theta_j+\varphi_2).
\label{eq:sm_onsite_gauge}
\end{align}
The derivative enters with a positive sign. Both representations retain the entire drive, without a high-frequency or rotating-wave approximation. Since \(F(T)=F(0)\), their propagators obey
\begin{equation}
U_{\rm ph}(T,0)=F(0)U_{\rm on}(T,0)F^\dagger(0).
\label{eq:sm_propagator_gauge}
\end{equation}
Here \(F(0)\ne I\), so the snapshot vectors must be transformed when comparing the two calculations.}

{\color{black}The pair exchanges used below are evaluated at \(B=20~{\rm G}=0.002~{\rm T}\), using the separation and polar angle of each pair. The finite 25-site device contains 300 pairs; the axial extension through \(L_w=641\) uses 1283 distinct pair-exchange calculations. The eigenproblem describes the effective spin Hamiltonian obtained from the approximate fixed-energy atomic reduction in Sec.~VI.}

{\color{black}The self-consistent BIC condition determines the bound branch. The observables defined above describe its spectral position, localization, chirality, and spatial decay.}

\section{General Floquet-Feshbach mechanism for a Hermitian Floquet BIC}
\label{sec:sm_feshbach_general}

The Floquet-Feshbach projection separates the condition for vanishing radiation from the eigenvalue equation for a bound state. The derivation applies to a general periodic Hamiltonian and does not require a Rydberg realization.

\subsection{From a periodic Hamiltonian to Sambe space}

Let the Hamiltonian be periodic,
\begin{equation}
H(t+T)=H(t),\qquad T=\frac{2\pi}{\Omega}.
\end{equation}
Floquet's theorem writes a solution of the time-dependent Schr\"odinger equation as
\begin{equation}
\ket{\psi(t)}=e^{-i\epsilon t}\ket{u(t)},\qquad \ket{u(t+T)}=\ket{u(t)}.
\end{equation}
Substitution gives
\begin{equation}
\left[H(t)-i\partial_t\right]\ket{u(t)}=\epsilon\ket{u(t)}.
\label{eq:sm_floquet_eigenproblem}
\end{equation}
The operator \(H(t)-i\partial_t\) acts on Sambe space,
\begin{equation}
{\cal H}_{S}={\cal H}_{\rm phys}\otimes{\cal T},
\end{equation}
where \({\cal T}\) is the space of periodic Fourier harmonics. We use the convention
\begin{equation}
H(t)=\sum_{r}H^{(r)}e^{-ir\Omega t},\qquad
\ket{u(t)}=\sum_m e^{-im\Omega t}\ket{u_m}.
\label{eq:sm_general_fourier}
\end{equation}
Equating the coefficients of each Fourier harmonic in Eq.~(\ref{eq:sm_floquet_eigenproblem}) gives
\begin{equation}
\sum_n \left[H^{(m-n)}-m\Omega\,\delta_{mn}\right]\ket{u_n}
=\epsilon\ket{u_m}.
\end{equation}
Thus
\begin{equation}
(H_F)_{mn}=H^{(m-n)}-m\Omega\,\delta_{mn}.
\end{equation}
The quasienergy representation has a gauge redundancy:
\begin{equation}
\epsilon\rightarrow \epsilon+q\Omega,\qquad m\rightarrow m-q.
\end{equation}
Only gauge-invariant statements, such as the set of open physical sideband energies and the normalized Floquet state over one period, carry physical meaning.

\subsection{Projectors and exact elimination of the continuum}

Suppose the physical Hilbert space is divided into a localized sector \(P\) and a continuum sector \(Q\), with \(P+Q=1\) and \(PQ=0\). In Sambe space these projectors lift to
\begin{equation}
{\cal P}=P\otimes I_T,\qquad {\cal Q}=Q\otimes I_T,
\end{equation}
because the separation into localized and continuum sectors acts on the physical coordinate without changing the Fourier label. With
\begin{equation}
\ket{u}=\ket{u_P}+\ket{u_Q},\qquad
\ket{u_P}={\cal P}\ket{u},\qquad
\ket{u_Q}={\cal Q}\ket{u},
\end{equation}
the Sambe eigenproblem becomes
\begin{align}
H_{F,PP}\ket{u_P}+H_{F,PQ}\ket{u_Q}
&=\epsilon\ket{u_P},\\
H_{F,QP}\ket{u_P}+H_{F,QQ}\ket{u_Q}
&=\epsilon\ket{u_Q}.
\end{align}
Eliminating the continuum with outgoing boundary conditions gives
\begin{equation}
\ket{u_Q}
=
\frac{1}{\epsilon+i0^+-H_{F,QQ}}H_{F,QP}\ket{u_P},
\end{equation}
and the reduced localized-sector equation
\begin{equation}
\left[
H_{F,PP}+\Sigma_F^R(\epsilon)
\right]\ket{u_P}
=\epsilon\ket{u_P},
\end{equation}
where
\begin{equation}
\Sigma_F^R(\epsilon)=
H_{F,PQ}
\frac{1}{\epsilon+i0^+-H_{F,QQ}}
H_{F,QP}.
\label{eq:sm_retarded_self_energy}
\end{equation}
The full time-periodic Hamiltonian remains Hermitian. The reduced operator becomes non-Hermitian only after the continuum is eliminated with retarded, outgoing-wave boundary conditions.

\subsection{Open-channel matrix and decay operator}

For the channel representation below, the continuum block \(QH(t)Q=H_w\) is time independent, as in the present system. If the continuum itself is driven, the same Feshbach construction must instead use its Floquet scattering modes.

Let \(\ket{k,\alpha;m}\) be a continuum eigenstate in physical band \(\alpha\) and Floquet sideband \(m\), with
\begin{equation}
H_{F,QQ}\ket{k,\alpha;m}
=
\left[E_\alpha(k)-m\Omega\right]\ket{k,\alpha;m}.
\end{equation}
At quasienergy \(\epsilon\), the physical sideband energy is
\begin{equation}
E_m=\epsilon+m\Omega.
\end{equation}
We use \(E_m=\epsilon+m\Omega\) throughout.
A channel \(c=(\alpha,m)\) is open only if \(E_m\) lies inside a continuum band and admits a real wave number. The spectral resolution of Eq.~(\ref{eq:sm_retarded_self_energy}) gives
\begin{equation}
\Sigma_F^R(\epsilon)=\Delta_F(\epsilon)-\frac{i}{2}\Gamma_F(\epsilon).
\end{equation}
For each independent on-shell solution, including distinct propagation directions at the same energy, define the weighted coupling vector
\begin{equation}
\ket{w_c(\epsilon)}
=
\sqrt{2\pi\rho_c(\epsilon)}\,\ket{v_c(\epsilon)},
\end{equation}
where \(\rho_c\) is the open-channel density of states and \(\ket{v_c}\) is the localized-sector coupling vector to that on-shell channel. Collecting these columns gives
\begin{equation}
W_{\rm open}(\epsilon)
=
\left(\ket{w_1},\ket{w_2},\ldots,\ket{w_{N_{\rm open}}}\right),
\end{equation}
and the decay operator factorizes as
\begin{equation}
\Gamma_F(\epsilon)
=
W_{\rm open}(\epsilon)W_{\rm open}^{\dagger}(\epsilon).
\label{eq:sm_gamma_factorization}
\end{equation}
For any localized Sambe vector \(\ket{a}\),
\begin{equation}
\bra{a}\Gamma_F\ket{a}
=
\|W_{\rm open}^{\dagger}\ket{a}\|^2
=
\sum_{c\in{\cal O}}2\pi\rho_c |\braket{v_c}{a}|^2\ge0.
\end{equation}
Therefore
\begin{equation}
\bra{a}\Gamma_F\ket{a}=0
\quad\Longleftrightarrow\quad
W_{\rm open}^{\dagger}\ket{a}=0.
\label{eq:sm_dark_condition}
\end{equation}
The decay rate is a sum of nonnegative contributions from independent channels. Destructive interference must therefore occur within each channel amplitude, and every on-shell radiation amplitude must vanish separately.

\subsection{Exact BIC criterion}

{\color{black}The following criterion requires the retarded inverse to be defined on the source and to yield a square-summable continuum component. The vanishing on-shell amplitude must cancel the resolvent singularity with sufficient regularity; threshold energies require separate analysis. The term real reduced operator refers to the Hermitian dispersive part, whose matrix entries need not be real.}

For real quasienergy, define the Hermitian reduced operator,
\begin{equation}
H_R(\epsilon)=H_{F,PP}+\Delta_F(\epsilon).
\end{equation}
\Needspace{8\baselineskip}
A Floquet BIC must satisfy both the radiation condition and the self-consistent eigenvalue equation at \(\epsilon_B\),
\begin{equation}
W_{\rm open}^{\dagger}(\epsilon_B)\ket{a_B}=0,
\label{eq:sm_bic_dark}
\end{equation}
and
\begin{equation}
H_R(\epsilon_B)\ket{a_B}=\epsilon_B\ket{a_B}.
\label{eq:sm_bic_self_consistent}
\end{equation}
Equivalently,
\begin{equation}
N_{\rm BIC}(\epsilon)
=
\dim\left[
\ker W_{\rm open}^{\dagger}(\epsilon)
\cap
\ker\left(H_R(\epsilon)-\epsilon I\right)
\right].
\label{eq:sm_bic_count}
\end{equation}
The rank-nullity theorem gives the number of radiation-dark directions,
\begin{equation}
D_{\rm dark}=N_P-\operatorname{rank}W_{\rm open},
\end{equation}
where \(N_P\) is the dimension of the localized Sambe manifold under consideration. This is a count of dark directions, not a count of BIC roots. A useful sufficient condition is that the Hermitian reduced operator preserve the dark subspace. If \(P_D\) projects onto \(\ker W_{\rm open}^{\dagger}\) and \(P_B=1-P_D\) projects onto the bright complement, then
\begin{equation}
P_B H_R P_D=0
\end{equation}
prevents \(H_R\) from rotating a dark vector into a bright direction.

\subsection{Rank of the channel-coupling matrix}

{\color{black}An example with four angular components illustrates why the number of open sidebands need not equal the rank of their coupling matrix. For a static contact column of angular label \(\ell_s=0\), multiplication by \(e^{-ir\theta_j}\) produces a column of label \(-r\bmod4\). Columns with harmonic orders differing by four can be parallel in this angular space. Adding such an open sideband increases the number of channels but need not increase \(\operatorname{rank}W_{\rm open}\). These columns specify the allowed contact couplings in the four-dimensional angular space. The state populations and joint symmetry require the full Floquet eigenvector, including its harmonic index (Sec.~III).}

A sideband whose coupling vector is parallel to an existing column leaves the rank unchanged. The resulting dark subspace contains a BIC only when a vector in it also solves the self-consistent real eigenvalue equation.

\subsection{Connection to the Rydberg construction}

{\color{black}For the axial effective Hamiltonian, the conserved quantity is the eigenvalue of a plaquette rotation combined with a quarter-period time translation. Its label is \(\kappa=m-\ell\bmod4\), where \(m\) is the state Fourier harmonic, not the order of an independently chosen phase-factor coefficient. The branch considered here has \(\kappa=2\) in the folded representative specified in Sec.~III. The guide component has nonzero amplitudes only at \(m=2\bmod4\). At the branch energy these guide replicas are outside the guide band, whereas replicas in other sectors are open. Radiation is excluded by a gap in the branch's symmetry sector within the global continuum, rather than by cancellation between propagating channels in that sector.}

{\color{black}In this case the on-shell condition in Eq.~(\ref{eq:sm_bic_dark}) follows from the absence of propagating guide channels in the state's symmetry sector. The bound state must also satisfy Eq.~(\ref{eq:sm_bic_self_consistent}). Section~IV relates this branch to a discrete eigenvalue in the gap of its symmetry sector. Section~V derives the tail of this branch from its contact source and long-range guide propagation. Both remain present when the propagating channels are closed.}

\section{Joint Floquet symmetry and plaquette chirality}
\label{sec:sm_sambe_bookkeeping}

The four plaquette sites are labeled by \(j=0,1,2,3\), with angular positions \(\theta_j=\pi j/2\). The angular-momentum basis is
\begin{equation}
b_\ell^\dagger=\frac{1}{2}\sum_{j=0}^{3}e^{i\ell\theta_j}p_j^\dagger,\qquad
p_j^\dagger=\frac{1}{2}\sum_{\ell=0,\pm1,2}e^{-i\ell\theta_j}b_\ell^\dagger .
\label{eq:sm_basis}
\end{equation}
The normalization follows from the discrete \(C_4\) orthogonality relation
\begin{equation}
\frac{1}{4}\sum_{j=0}^{3}e^{i(\ell-\ell')\theta_j}
=\delta_{\ell,\ell'}^{(4)},
\label{eq:sm_c4_orthogonality}
\end{equation}
where \(\delta_{\ell,\ell'}^{(4)}=1\) when \(\ell-\ell'\) is a multiple of four and is zero otherwise. This is why all angular labels in the four-site plaquette are defined modulo four. In particular, \(\ell=3\) and \(\ell=-1\) describe the same \(C_4\) representation.

{\color{black}For the complete circulant plaquette matrix, write \(K_{j,j+d}=k_d\), with \(k_0=V_0\), \(k_3=k_1^*\), and real \(k_2\). Applying \(K\) to the angular ket gives
\begin{equation}
E_\ell^{(0)}
=\sum_{d=0}^3 k_d e^{i\ell\pi d/2}
=V_0+2\operatorname{Re}(k_1e^{i\ell\pi/2})+k_2(-1)^\ell .
\label{eq:sm_bare_plaquette}
\end{equation}
Only the special choice \(k_1=-K_{\rm nn}\), \(k_2=0\) gives the nearest-neighbor expression \(V_0-2K_{\rm nn}\cos(\ell\pi/2)\). The real axial zero-regulator matrix has \(E_{+1}^{(0)}=E_{-1}^{(0)}\), but the driven eigenstate is determined by the full angular-harmonic problem, not by these bare levels alone.}

\subsection{\texorpdfstring{{\color{black}Joint rotation and time translation}}{Joint rotation and time translation}}

{\color{black}Let \(R\ket{p_j}=\ket{p_{j+1\bmod4}}\) and \(R\ket{s_n}=\ket{s_n}\). Exact axial symmetry requires \(RH_0R^\dagger=H_0\), so that \(K\) is circulant and every column of \(C\) is constant around the plaquette. The waveform obeys \(\phi_{j-1}(t)=\phi_j(t+T/4)\), including its doubled winding in the second harmonic. Therefore
\begin{equation}
RH_g(t)R^\dagger=H_g(t+T/4),\qquad
(\mathcal G u)(t)=Ru(t-T/4),\qquad
[\mathcal G,H_{F,g}]=0,
\label{eq:sm_joint_covariance}
\end{equation}
where \(g={\rm ph},{\rm on}\) denotes the representation. The instantaneous Hamiltonian need not commute with \(R\). Since \(R\ket{\ell}=e^{-i\ell\pi/2}\ket{\ell}\),
\begin{equation}
\mathcal G\bigl(\ket{\ell}e^{-im\Omega t}\bigr)
=e^{i(m-\ell)\pi/2}\ket{\ell}e^{-im\Omega t},
\qquad \kappa=m-\ell\pmod4 .
\label{eq:sm_joint_sector}
\end{equation}
The transformation \(F\) commutes with \(\mathcal G\), so both representations have the same sector label. A guide site is invariant under \(R\) and hence has harmonics only at \(m=\kappa\bmod4\).}

{\color{black}The onsite representation contains only the two imposed harmonic orders. For \(h=1,2\), use amplitude \(z_h\) and phase \(\varphi_h\) from Eq.~(\ref{eq:sm_drive_phase}):
\begin{equation}
W_{+h}=H_{\rm on}^{(+h)}
=\frac{\Omega z_h}{2}\sum_j
e^{ih\theta_j-i\varphi_h}\ket{p_j}\bra{p_j},
\qquad W_{-h}=W_{+h}^\dagger .
\label{eq:sm_onsite_harmonics}
\end{equation}
Using the angular orthogonality relation,
\begin{equation}
\bra{\ell'}\operatorname{diag}(e^{ih\theta_j})\ket{\ell}
=\frac14\sum_j e^{i(\ell+h-\ell')\theta_j}
=\delta_{\ell',\ell+h}^{(4)} .
\label{eq:sm_harmonic_delta}
\end{equation}
Thus \(W_{+h}\) connects \((\ell,m)\) to \((\ell+h,m+h)\); \(W_{-h}\) changes both labels by \(-h\). Each step preserves \(m-\ell\). The only onsite-frame harmonic couplings are \(h=\pm1,\pm2\), even though the phase representation contains infinitely many harmonics.}

{\color{black}Write \(u_{\rm on}(t)=\sum_m b_m e^{-im\Omega t}\), \(b_{\ell m}=\braket{\ell}{b_m}\), and \(q_{n,m}=\braket{s_n}{b_m}\). Within a sector define \(\ell_m=m-\kappa\bmod4\) and \(\beta_m=b_{\ell_m m}\). The contact vector to the angular mode \(\ell=0\) is \(v_n=\bra{s_n}H_0\ket{\ell=0}\). If \(C_{jn}=c_n^{(0)}\) is independent of \(j\), then \(v_n=2c_n^{(0)*}\). The exact sector equations are
\begin{align}
\epsilon\beta_m
&=(E_{\ell_m}^{(0)}-m\Omega)\beta_m
+\sum_{h=1}^2\frac{\Omega z_h}{2}
 \left(e^{-i\varphi_h}\beta_{m-h}+e^{i\varphi_h}\beta_{m+h}\right)
+\mathbf1_{m=\kappa\ (4)}\,v^\dagger q_m,\\
(\epsilon+m\Omega-H_w)q_m&=v\beta_m,\qquad m=\kappa\pmod4 .
\label{eq:sm_sector_equations}
\end{align}
No guide coordinates occur at forbidden \(m\) in this exact sector. These equations retain all plaquette links and guide couplings.}

\begin{figure}[t]
\includegraphics[width=0.90\textwidth]{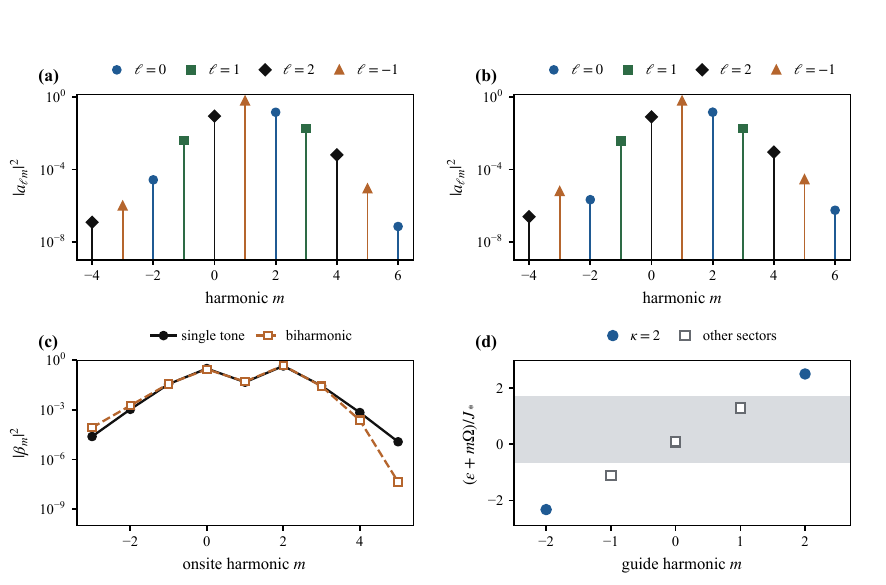}
\caption{{\color{black}Angular-harmonic structure and accessible guide replicas of the same axial Floquet state. (a,b) Phase-frame weights \(|a_{\ell m}|^2\) for the single-tone and biharmonic modes over \(-4\le m\le6\); the displayed coordinates obey \(\kappa=m-\ell=2\bmod4\). These are not Fourier weights of the imposed phase factor. (c) Onsite-frame weights \(|\beta_m|^2\) for both waveforms over \(-3\le m\le5\). (d) Single-tone sideband energies \((\epsilon+m\Omega)/J_*\) and the guide band. Filled circles denote the sector-allowed guide harmonics \(m=2\bmod4\); hollow squares denote other sectors. Each mode is normalized in the full single-excitation space. Parameters are direct \(20~{\rm G}\) effective couplings, \(L_w=321\), \(\Omega/J_*=1.21\), \(z_1=1.1\), \(\varphi_1=0.49\pi\), and \(\varphi_2=0\), with waveform endpoints \(z_2=0,0.1\). The onsite cutoff is \(M=16\); phase-frame coefficients follow from coherent periodic reconstruction.}}
\label{fig:sm_sambe_channels}
\end{figure}

{\color{black}For each waveform, the Floquet eigenstate studied here belongs to \(\kappa=2\) in the chosen folded representative, with \(\epsilon/J_*\simeq0.09057\) and \(0.1215\) for the single-tone and biharmonic drives, respectively. The same branch is followed by overlap as the guide length and the second-harmonic amplitude are varied (Sec.~VIII). Symmetry classifies this branch; it does not uniquely choose its sector or rule out other BICs. The phase-frame angular-harmonic populations and guide-channel energies are shown in Fig.~\ref{fig:sm_sambe_channels}. The \(m=\pm2\) guide channels are closed in energy; the globally open \(m=0,1\) replicas belong to other sectors. Changing the quasienergy representative requires the simultaneous relabeling
\begin{equation}
\epsilon'=\epsilon+q\Omega,\qquad m'=m-q,\qquad
\kappa'=\kappa-q\pmod4.
\label{eq:sm_sector_relabel}
\end{equation}
This leaves \(\epsilon'+m'\Omega=\epsilon+m\Omega\) invariant. A sector label quoted without its quasienergy representative cannot determine which physical channels are open.}

\Needspace{6\baselineskip}
\subsection{\texorpdfstring{{\color{black}Hybridization of Floquet components}}{Hybridization of Floquet components}}

{\color{black}Eliminating the closed guide blocks in Eq.~(\ref{eq:sm_sector_equations}) gives the energy-dependent plaquette-harmonic matrix
\begin{align}
\mathcal K_{mn}(\epsilon)
&=(K_F^{P})_{mn}+\delta_{mn}\Sigma_m(\epsilon),\\
\Sigma_m(\epsilon)
&=\mathbf1_{m=\kappa\ (4)}\,
v^\dagger(\epsilon+m\Omega-H_w)^{-1}v ,
\label{eq:sm_guide_selfenergy}
\end{align}
where \(K_F^P\) is the plaquette part, including the onsite Fourier couplings and \(-m\Omega\). This elimination is exact for the finite Sambe matrix whenever the inverse exists. The self-energy is fixed by the guide Hamiltonian and the quasienergy.}

{\color{black}The block containing the three components \(m=0,1,2\) can be displayed while retaining its coupling to all other harmonics. Let \(A=\{m=0,1,2\}\) and let \(B\) denote the remaining plaquette harmonics in \(\mathcal K\). Partition \(\beta=(\beta_A,\beta_B)\). The second block equation yields
\begin{equation}
\beta_B=(\epsilon I_B-\mathcal K_{BB})^{-1}
\mathcal K_{BA}\beta_A ,
\end{equation}
and substitution into the first gives
\begin{equation}
\left[\mathcal K_{AA}
+\mathcal K_{AB}(\epsilon I_B-\mathcal K_{BB})^{-1}
\mathcal K_{BA}\right]\beta_A=\epsilon\beta_A .
\label{eq:sm_three_coordinate}
\end{equation}
All blocks on the left are evaluated at the same \(\epsilon\). The reduced equation includes the energy-dependent contribution of every eliminated harmonic and remains equivalent to the full finite eigenproblem.}

{\color{black}The single-tone \(L_w=321\) mode has bare diagonal energies \(0.5731J_*\), \(-0.2478J_*\), and \(-0.6443J_*\) at \(m=0,1,2\), respectively. The \(m=2\) guide self-energy is \(+0.3530J_*\), while each fundamental drive link has magnitude \(\Omega z_1/2=0.6655J_*\). Their finite detunings and coherent coupling produce onsite weights \(0.3142\), \(0.04699\), and \(0.4499\), respectively. The three retained coordinates carry weight \(0.8112\); all onsite plaquette harmonics together carry period-averaged weight \(0.8772\). The bare \(m=1,\ell=-1\) weight alone is therefore insufficient to explain the larger phase-frame chirality, which includes interference from the transformation of the full state.}

\Needspace{6\baselineskip}
\subsection{\texorpdfstring{{\color{black}Phase coefficients and coherent chirality}}{Phase coefficients and coherent chirality}}

{\color{black}To keep phase-factor order distinct from state harmonic \(m\), denote it by \(r\):
\begin{equation}
f_j(t)\equiv e^{i\phi_j(t)}
=\sum_r c_r e^{ir\Omega t-ir\theta_j}.
\label{eq:sm_winding}
\end{equation}
Here \(f_j(t)\) is a dimensionless local phase factor; the guide source \(f_n(t)\) introduced in Sec.~V has a guide-site index and angular-frequency units. Expanding the two sine phases and multiplying their series gives
\begin{align}
c_r
&=\sum_{v\in\mathbb Z}J_{r-2v}(z_1)J_v(z_2/2)
 e^{i(r-2v)\varphi_1+iv\varphi_2}\\
&=e^{ir\varphi_1}\sum_vJ_{r-2v}(z_1)J_v(z_2/2)
 e^{iv(\varphi_2-2\varphi_1)} .
\label{eq:sm_phase_coefficients}
\end{align}
The Bessel functions \(J_r\) are of the first kind. At \(z_2=0\), \(c_r=J_r(z_1)e^{ir\varphi_1}\). The term with order \(r\) has time factor \(e^{ir\Omega t}\), so it belongs to Fourier component \(-r\) under Eq.~(\ref{eq:sm_general_fourier}).}

{\color{black}For consistency with the static-channel notation, define
\begin{equation}
J_{\ell_s n}^{(0)}
=\frac12\sum_j e^{-i\ell_s\theta_j}J_{jn}^{(0)},\qquad
J_{jn}^{(0)}=C_{jn}.
\label{eq:sm_static_channel}
\end{equation}
Multiplying this contact column by \(e^{-ir\theta_j}\) gives
\begin{equation}
\ell_{\rm eff}=\ell_s-r\pmod4.
\label{eq:sm_sambe_product}
\end{equation}
This relation fixes the angular label of a contact harmonic. It does not determine the Floquet-state populations or establish the absence of radiation. In particular, \(c_1\) acting on an axial static column can have angular label \(-1\), but the populated guide replicas of the solved state are still fixed by \(\kappa=2\).}

{\color{black}Let \(u_{\rm ph}(t)=\sum_m a_m e^{-im\Omega t}=F(t)u_{\rm on}(t)\). Multiplying the Fourier series at a plaquette site and then projecting onto \(\ket{\ell}\) yields
\begin{align}
a_{j,m}&=\sum_r c_r e^{-ir\theta_j}b_{j,m+r},\\
a_{\ell m}&=\sum_r c_r b_{\ell+r,m+r},\qquad
\braket{s_n}{a_m}=q_{n,m}.
\label{eq:sm_coherent_transform}
\end{align}
Both the harmonic and angular labels shift by \(r\), so every term retains \(m-\ell\). For the dominant phase-frame coefficient,
\begin{equation}
a_{-1,1}=\sum_r c_r b_{-1+r,1+r}.
\label{eq:sm_chiral_component}
\end{equation}
The terms interfere as complex amplitudes, so their squared magnitudes cannot be treated as independent probabilities. The coherent sum and the direct Fourier transform of the periodic state give the same coefficient to numerical precision.}

{\color{black}The full target-chirality snapshot and its period average are different quantities:
\begin{align}
P_{\chi_-}(t_0)
&=\left|\sum_m a_{-1,m}e^{-im\Omega t_0}\right|^2,\\
\overline P_{\chi_-}
&=\frac1T\int_0^T P_{\chi_-}(t)\,dt
=\sum_m|a_{-1,m}|^2 .
\label{eq:sm_chirality_snapshot_average}
\end{align}
The integral removes cross terms with unequal \(m\); the snapshot retains them. At \(t_0=0\), the single-tone values are \(P_{\chi_-}=0.6151\) and \(\overline P_{\chi_-}=0.6182\), while the biharmonic values are \(0.6211\) and \(0.6255\). The chirality projector must also transform: in the onsite representation the same observable is \(F^\dagger(t)\ket{-1}\bra{-1}F(t)\). Applying an untransformed angular projector to \(b_m\) measures a different quantity. Site populations and snapshot IPR are unchanged by the local phase transformation.}

{\color{black}The dominant contributions to this coherent sum can be identified explicitly. For the single-tone \(L_w=321\), \(M=16\) state, the contributions to \(a_{-1,1}\) from onsite harmonics \(m=0,1,2\) are \(c_{-1}b_{2,0}\), \(c_0b_{-1,1}\), and \(c_1b_{0,2}\), respectively, with angular labels understood modulo four. Their magnitudes are \(0.2640\), \(0.1560\), and \(0.3159\), and they have the same phase: a common overall rephasing makes all three positive. Including the remaining onsite harmonics in the coherent sum gives \(|a_{-1,1}|^2=0.61817476\), close to \(\overline P_{\chi_-}=0.61818436\); the other phase-frame harmonics contribute only \(9.60\times10^{-6}\) to that average. The large chirality therefore follows from constructive interference among hybridized onsite components, not from the onsite \(m=1\) population alone.}

{\color{black}Figure~\ref{fig:sm_waveform} separates the input coefficients from the state response. The waveform family contains the five choices \(z_2=0,0.025,0.05,0.075,0.1\); geometry, static couplings, onsite terms, \(J_*\), \(\Omega\), \(z_1\), and both phases are fixed. Adjacent normalized snapshots have squared overlap exceeding \(0.99996\). The large overlap is consistent with continuity within this symmetry-preserving family. Responses to other waveforms, disorder, and open-system loss require separate analysis.}

\section{Localized Floquet branch and extended states}
\label{sec:sm_closed_diagnostics}

{\color{black}We identify the finite branch from the closed one-period propagator and follow it by overlap as the guide length is increased. For a normalized state \(u\), let \(P_{\rm plaq}\) be the plaquette weight, \(P_{\chi_-}\) the physical phase-gauge weight in the target chirality at \(t=0\), IPR the inverse participation ratio, and \(\Delta_{\rm ext}\) the distance to the nearest waveguide-like extended state. The first three refer to the full normalized state at a fixed time, without conditioning on plaquette occupation or averaging over a period.}

{\color{black}For the tracked mode \(B\), define}
\[
\mathcal E_{\rm ext}
=\{\nu\ne B:\ P_{{\rm plaq},\nu}(0)<0.15\}.
\]
\begin{equation}
\Delta_{\rm ext}
=\min_{\nu\in\mathcal E_{\rm ext}}
 \min_{q\in\mathbb Z}
 |\epsilon_B-\epsilon_\nu+q\Omega|.
\label{eq:sm_extended_distance}
\end{equation}
{\color{black}The distance has the same angular-frequency units as the quasienergies; Fig.~2(a) shows \(\Delta_{\rm ext}/J_*\). Changing the cutoff to 0.10 or 0.20 selects the same nearest level for the axial size sequence. This distance measures the separation of finite-chain eigenvalues, rather than a decay linewidth.}

\subsection{\texorpdfstring{{\color{black}Sector gap and the folded continuum}}{Sector gap and the folded continuum}}
\label{sec:sm_infinite_certificate}

{\color{black}The combined rotation and quarter-period time translation decomposes the Floquet-Sambe operator into sectors labeled by \(\kappa=m-\ell\pmod4\). In this subsection \(\mathcal B\), \(\epsilon\), and \(\Omega\) are restored to common angular-frequency units; numerical intervals below are divided by \(J_*\). If \(\mathcal B\) is the full-line guide band, the essential spectrum of a sector is the union of the guide replicas accessible in that sector,}
\begin{equation}
\sigma_{\rm ess}\!\left(H_F^{(\kappa)}\right)
 =\bigcup_{m=\kappa\;({\rm mod}\,4)}\left(\mathcal B-m\Omega\right).
\label{eq:sm_sector_essential_spectrum}
\end{equation}
{\color{black}The tracked branch has \(\kappa=2\), so its allowed guide harmonics satisfy \(m=2\pmod4\). At the branch quasienergy, the nearest allowed physical energies \(E_{\pm2}=\epsilon\pm2\Omega\) lie outside the guide band \([-0.6223,1.6917]J_*\). The in-band energies \(E_0\) and \(E_1\) occur in other sectors, supplying the global folded continuum while the branch remains in its own sector gap.}

The distance to the continuum in the same sector is
\begin{equation}
 d_\kappa(\epsilon)=\operatorname{dist}\!\left(\epsilon,\sigma_{\rm ess}\!\left(H_F^{(\kappa)}\right)\right).
\label{eq:sm_sector_gap}
\end{equation}

{\color{black}For the real scalar half-line guide, extending a vector by zero to the full line bounds its quadratic form by the range of the continuous symbol}
\[
E_w(k)=\omega_w+2\sum_{d\ge1}t_d\cos(dk).
\]
{\color{black}Conversely, truncated plane waves translated away from the boundary form approximate eigenstates whose residual tends to zero at each energy in the dispersion (Weyl sequences). The spectrum is therefore the interval covered by the dispersion and is entirely essential. For the specified guide spacing and directly evaluated hopping through \(d=1024\), the band is bounded by}
\begin{equation}
[-0.6222518,1.6916892]\subseteq \mathcal B/J_*
\subseteq[-0.6222530,1.6916899].
\label{eq:sm_guide_band_enclosure}
\end{equation}
{\color{black}The comparison guide has the same onsite energy \(0.70J_*\) and hopping \(J_*j_\infty/d^3\), with dimensionless \(j_\infty=w_g/(a_w^3\nu_*)\). Its operator differs from the pair-derived guide by less than \(3.619\times10^{-7}J_*\) in norm. Here \(w_g=-7.459068588401653~{\rm GHz}\,\mu{\rm m}^3\) is the bare axial dipolar exchange coefficient between the ordered target pair states; the displayed digits specify the fixed numerical input. The comparison-band endpoints, divided by \(J_*\), are \(0.70+2j_\infty\zeta(3)\) and \(0.70-(3/2)j_\infty\zeta(3)\). The inner interval in Eq.~(\ref{eq:sm_guide_band_enclosure}) places the state inside the continuum, whereas the outer interval places its nearest allowed guide energies outside the band.}

\subsection{Existence of the embedded state in the infinite guide}

{\color{black}We use the onsite representation, in which the contact and guide couplings are time independent. The half-line guide has bounded, absolutely summable hopping, and each plaquette--guide contact column is square summable. The Floquet operator is self-adjoint on the domain}
\[
\mathcal D=
\left\{u:\sum_m m^2\|u_m\|^2<\infty\right\}.
\]
{\color{black}Within a fixed joint sector, the uncoupled plaquette harmonic ladder has compact resolvent. The contact and finite-harmonic plaquette terms are compact relative to the uncoupled resolvent: a finite input-harmonic restriction has finite rank, and the omitted-harmonic norm tends to zero. Consequently the essential spectrum is Eq.~(\ref{eq:sm_sector_essential_spectrum}). For an \(L,M\) Floquet state extended by zero, the residual has separate contributions from the retained coordinates, omitted harmonics, and omitted guide sites. Let \(u_{L,M}\) be the normalized Floquet-Sambe vector obtained with guide length \(L\) and harmonic cutoff \(|m|\le M\), and extend it by zero to the omitted guide sites and harmonics. If \(\mathcal R_{L,M}=H_F^{(\kappa)}-\epsilon_{L,M}\), the omitted components have disjoint supports, so}
\begin{equation}
\left\|\mathcal R_{L,M}u_{L,M}\right\|^2
=\rho_{\rm kept}^2+\rho_{\rm harm}^2+\rho_{\rm space}^2 .
\label{eq:sm_residual_decomposition}
\end{equation}
{\color{black}Write \(b_{P,m}\) for the four-component plaquette coordinate and \(q_{r,m}\) for the guide coordinate. The finite vector is zero outside its retained support. With \(W_s\) the onsite Fourier operators of Eq.~(\ref{eq:sm_onsite_harmonics}), the residual components outside the retained space are}
\begin{align}
e^P_h&=\sum_{s=\pm1,\pm2}W_s b_{P,h-s},
 && |h|>M,\\
e^Q_{n,m}&=\sum_{j=0}^{3}C_{jn}^{*}b_{j,m}
 +\sum_{r=1}^{L}t_{n-r}q_{r,m},
 && n>L,\quad |m|\le M.
\label{eq:sm_omitted_components}
\end{align}
{\color{black}Only \(M<|h|\le M+2\) can be nonzero in the first line. Because the contact and guide terms enter the same guide component, they are summed as complex amplitudes before a modulus or norm is taken in the explicitly evaluated region outside the retained guide. The retained residual, omitted harmonics, and omitted spatial components are supported on orthogonal coordinates.}
{\color{black}For the pair-derived guide, the long-distance matrix elements are bounded from the same fixed-energy atomic reduction used to construct the operator. For \(n>N>L\), use}
\begin{equation}
\frac{|C_{jn}|}{J_*}\le c_j(n-\rho_c)^{-3},\qquad
\frac{|t_{n-r}|}{J_*}\le c_g(n-r-\beta_g)^{-3},
\label{eq:sm_shifted_inverse_cube_bounds}
\end{equation}
{\color{black}with \(N>\max(\rho_c,L+\beta_g)\), and with the bounds uniform in \(r\le L\). In the axial sector the first sum reduces to \(2c_c|\beta_m|\), where \(c_c\) is the common contact bound. Minkowski's inequality and the integral estimate for the square tail give}
\begin{equation}
\left(\sum_{n>N}|e^Q_{n,m}/J_*|^2\right)^{1/2}
\le
\frac{\sum_j c_j|b_{j,m}|}
 {\sqrt5\,(N-\rho_c)^{5/2}}
\;+\;
\frac{c_g\sum_{r=1}^{L}|q_{r,m}|}
 {\sqrt5\,(N-L-\beta_g)^{5/2}}.
\label{eq:sm_shifted_tail_bound}
\end{equation}
{\color{black}Because the guide contribution depends on \(n-r\), its tail bound retains the distance \(N-L-\beta_g\). In cyclic-frequency units, the dipole-only pair matrix in either fixed atomic basis is}
\begin{equation}
A(R,\theta)=D+R^{-3}W(\theta),
\qquad
\delta_Q=\operatorname{dist}\!\left(E_0,\sigma(D_{QQ})\right).
\label{eq:sm_pair_operator_definition}
\end{equation}
{\color{black}Here \(D\) is independent of separation, \(W\) has units \({\rm GHz}\,\mu{\rm m}^3\), and the subscripts refer to the atomic target subspace and its complement. For the target/complement partition used here, the static block \(D_{PQ}\) vanishes. Factoring out \(E_0-D_{QQ}\) gives, whenever \(\delta_Q>\|W_{QQ}(\theta)\|/R^3\),}
\begin{equation}
\left\|(E_0-A_{QQ})^{-1}\right\|
\le\left[\delta_Q-\frac{\|W_{QQ}(\theta)\|}{R^3}\right]^{-1}.
\label{eq:sm_pair_inverse_bound}
\end{equation}
{\color{black}The constants used for the two \(L=641\) states are \(c_c=10.10173898\), \(c_g=0.5499958290\), \(\rho_c=1\), and \(\beta_g=(0.500001\times10^{-6}\,\mu{\rm m})/a_w\). They follow from the atomic bounds \(W_c\le137~{\rm GHz}\,\mu{\rm m}^3\), \(W_{QQ}\le23.404305~{\rm GHz}\,\mu{\rm m}^3\), \(W_{PQ}\le5.216655~{\rm GHz}\,\mu{\rm m}^3\), \(\delta_c\ge3.4037\times10^{-5}~{\rm GHz}\), and \(\delta_g\ge0.0671738472~{\rm GHz}\), with directly evaluated pair entries taken in the zero-regulator limit. The fixed-energy reduction obeys, whenever the denominator is positive,}
\begin{equation}
\left|J^{\rm cyc}(R,\theta)-\frac{W_{DP,PD}(\theta)}{R^3}\right|
\le
\frac{\|W_{PQ}(\theta)\|^2}
 {R^6[\delta_Q-\|W_{QQ}(\theta)\|/R^3]}.
\label{eq:sm_pair_reduction_bound}
\end{equation}
{\color{black}The bound is applied only beyond the directly evaluated pair region; all retained finite entries remain pair-derived. The inverse-cube spatial tail has the square-tail estimate}
\begin{equation}
\sum_{n>N}(n-\rho)^{-6}
\le \frac{1}{5(N-\rho)^5},
\qquad N>\rho .
\label{eq:sm_inverse_cube_residual_bound}
\end{equation}
{\color{black}Together with the contact and guide bounds in Eq.~(\ref{eq:sm_shifted_inverse_cube_bounds}), this inverse-cube square-tail estimate gives the analytic remainder used in Eq.~(\ref{eq:sm_shifted_tail_bound}) and Table~I, with the appropriate shifts \(\rho_c\) and \(L+\beta_g\). For any normalized trial vector, the elementary spectral inequality}
\begin{equation}
\operatorname{dist}\!\left(\epsilon_{L,M},\sigma\!\left(H_F^{(\kappa)}\right)\right)
\le \left\|\mathcal R_{L,M}u_{L,M}\right\|
\label{eq:sm_residual_spectrum}
\end{equation}
{\color{black}follows by expanding the vector in the spectral resolution of the Hermitian sector. Thus, when the residual is smaller than the distance to the sector essential spectrum, the interval \([\epsilon_{L,M}-\rho,\epsilon_{L,M}+\rho]\), with \(\rho=\lVert\mathcal R_{L,M}u_{L,M}\rVert\), remains inside the sector gap and contains a spectral point. That point is discrete in the sector gap, while the same quasienergy is embedded in the folded spectrum of the other sectors.}

{\color{black}The direct calculations at \(L_w=21,41,81,161,321,641\), with harmonic cutoffs \(M=8,12,16\), converge along the same branch for each waveform. Guide hoppings are evaluated directly through separation \(d=1024\). The contact couplings are evaluated directly through site 689 and at site 1024; the remaining 334 contact entries up to site 1024 are bounded using the pair-reduction estimate. At directly evaluated sites, contact and guide contributions are added coherently before taking a norm. The other sites enter through an upper bound on their contact contribution. For the \(L=641\), \(M=16\), \(N=1024\) trials, Table~I gives the residual bounds. The analytic tail uses \(N-L-\beta_g\), rather than \(N\), in the guide contribution. The corresponding same-sector essential-gap lower bounds are \(0.8188J_*\) and \(0.8498J_*\), and both residual intervals lie inside the continuum of another sector with margins greater than \(0.71J_*\).}
\begin{table}[!htb]
\caption{\color{black}Residual bounds for the zero-extended \(L=641\), \(M=16\) Floquet trials. All entries are norm bounds after division by \(J_*\). In the outer region, directly evaluated contact and guide amplitudes are added coherently, while the remaining contact entries are bounded analytically.}
\label{tab:sm_residual_budget}
\centering
\begin{tabular}{lcccccc}
\toprule
waveform \((L/M/N)\) & retained & harmonic & outer sites & analytic tail & matrix error & total\\
\midrule
single tone \((641/16/1024)\)
& \(2.16{\times}10^{-11}\) & \(5.23{\times}10^{-17}\)
& \(4.72{\times}10^{-7}\) & \(2.26{\times}10^{-7}\)
& \(3.37{\times}10^{-10}\) & \(5.24{\times}10^{-7}\)\\
biharmonic \((641/16/1024)\)
& \(2.24{\times}10^{-11}\) & \(1.81{\times}10^{-15}\)
& \(4.86{\times}10^{-7}\) & \(2.32{\times}10^{-7}\)
& \(3.37{\times}10^{-10}\) & \(5.39{\times}10^{-7}\)\\
\bottomrule
\end{tabular}
\end{table}
\FloatBarrier

{\color{black}The entries in the table are norm bounds, not squared norms. Their combination is}
\begin{equation}
\bar\rho=\left(\bar\rho_{\rm kept}^2+\bar\rho_{\rm harm}^2+\bar\rho_{\rm shell}^2+\bar\rho_{\rm tail}^2\right)^{1/2}+\delta_{\rm assembly}.
\label{eq:sm_total_residual_bound}
\end{equation}
{\color{black}The last term bounds the error in constructing the finite matrix. With \(\bar\rho\) given by the last column,}
\begin{equation}
I=[\epsilon_{L,M}-\bar\rho,\epsilon_{L,M}+\bar\rho],
\qquad
I\cap\sigma_{\rm ess}(H_F^{(2)})=\varnothing,\qquad
I\subset {\rm int}\,\mathcal B
\subset \sigma_{\rm ess}(H_F^{(0)}).
\label{eq:sm_embedding_interval}
\end{equation}
{\color{black}The \(m=0,\kappa=0\) guide replica therefore supplies the other-sector continuum containing the residual interval. For a normalized trial in the self-adjoint operator domain, Eq.~(\ref{eq:sm_residual_spectrum}) gives \(\operatorname{dist}(\epsilon_{L,M},\sigma(H_F^{(2)}))\le\bar\rho\). The interval \(I\) therefore contains a sector eigenvalue. Since \(I\) is separated from that sector's essential spectrum and lies inside the continuum of another sector, the eigenvalue has a normalizable eigenstate embedded in the continuum of the full effective operator. This argument gives existence without determining the number of embedded eigenstates.}

{\color{black}The sector gap forbids radiation at the state's energy, while the self-consistent eigenvalue equation and finite-size convergence identify the bound branch. Its chirality follows from the coherent mixing derived in Sec.~III. Neither the large chirality weight nor \(\Delta_{\rm ext}\) alone is a BIC criterion.}

\FloatBarrier
\section{Source and guide propagation in the algebraic tail}
\label{sec:sm_tail_feshbach}

{\color{black}The physical projectors \(P,Q\) in Eq.~(\ref{eq:sm_projectors}) lift to \(\mathcal P=P\otimes I_T\) and \(\mathcal Q=Q\otimes I_T\) in Sambe space. With \(u_P=\mathcal P u\) and \(u_Q=\mathcal Q u\), the projected equations are
\begin{align}
H_{F,PP}u_P+H_{F,PQ}u_Q&=\epsilon u_P,\\
H_{F,QP}u_P+H_{F,QQ}u_Q&=\epsilon u_Q.
\end{align}
Rearranging the second equation gives
\begin{equation}
(\epsilon-H_{F,QQ})u_Q=H_{F,QP}u_P .
\label{eq:sm_q_equation}
\end{equation}
Whenever the source belongs to the domain of the required inverse,
\begin{equation}
u_Q=G_Q(\epsilon)H_{F,QP}u_P,\qquad
G_Q(\epsilon)=(\epsilon-H_{F,QQ})^{-1}.
\label{eq:sm_q_solution}
\end{equation}
In the branch's exact \(\kappa=2\) sector, only guide harmonics \(m=2\bmod4\) are populated, and every corresponding \(\epsilon+m\Omega\) lies outside the guide band. The guide resolvent is therefore bounded on every populated block; open replicas in other sectors do not enter this inversion. Accessible propagating modes would instead require an outgoing boundary prescription and removal of on-shell components. The source and the resolvent both enter the spatial profile.}

\subsection{\texorpdfstring{{\color{black}Guide source and harmonic response}}{Guide source and harmonic response}}

{\color{black}At a fixed time the guide source of the normalized state is the vector
\begin{align}
f(t)&=QH_{\rm ph}(t)P u_{\rm ph}(t)\\
&=C^\dagger F_P^\dagger(t)F_P(t)P u_{\rm on}(t)
=C^\dagger P u_{\rm on}(t).
\label{eq:sm_actual_source}
\end{align}
The phase factors cancel in the second line because \(QF=Q\), while the driven state remains time dependent. This relation holds before imposing axial symmetry. The derivative part of \(H_F\) has no \(QP\) block because \(P\) and \(Q\) are time independent. Fourier transformation then gives
\begin{equation}
b_{P,m}\equiv P b_m,\qquad
f_m=C^\dagger b_{P,m},\qquad
(\epsilon+m\Omega-H_w)q_m=f_m.
\label{eq:sm_source_response}
\end{equation}
Both \(q_m\) and \(f_m\) are unchanged by the phase-to-onsite transformation. Using \(C^\dagger Pa_m\) in place of \(f_m\) would omit the phase-frame contact factor. Under exact axial symmetry, Eq.~(\ref{eq:sm_source_response}) reduces to \(f_m=v\beta_m\) in the allowed guide harmonics.}

{\color{black}For a uniform time grid \(t_l=lT/N_t\), with \(l=0,\ldots,N_t-1\), the Fourier convention gives
\begin{equation}
u_m=\frac1{N_t}\sum_{l=0}^{N_t-1}e^{+2\pi iml/N_t}u(t_l).
\label{eq:sm_discrete_fourier}
\end{equation}
The endpoint \(t=T\) is not counted twice. The number of time samples \(N_t\) controls aliasing, whereas the onsite cutoff \(M\) controls harmonic truncation. Time-domain, Fourier, and harmonic-equation reconstructions of the source and response agree at the \(10^{-15}\) level for all states considered. Both chirality measures and the guide profiles are evaluated from those same Floquet modes.}

\begin{figure}[t]
\includegraphics[width=0.90\textwidth]{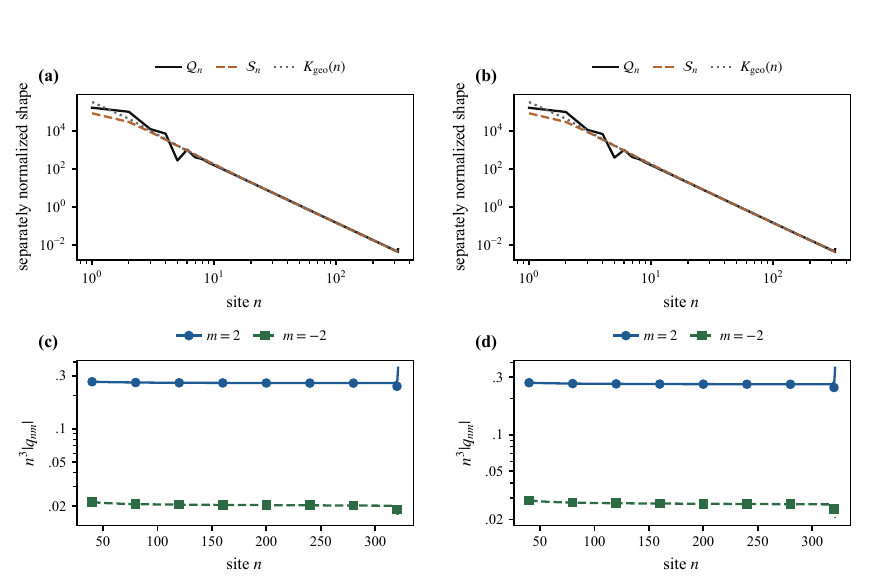}
\caption{{\color{black}Source and waveguide response of the same normalized axial Floquet mode. (a,b) Period-rms envelopes \(\mathcal S_n=(\sum_m|f_{n,m}|^2)^{1/2}\) and \(\mathcal Q_n=(\sum_m|q_{n,m}|^2)^{1/2}\), compared with \(K_{\rm geo}(n)\) for the single-tone and biharmonic states, respectively. Each curve is separately normalized at \(n_0=\max(12,\lfloor L_w/6\rfloor)=53\). (c,d) Unnormalized \(n^3|q_{n,m}|\) for \(m=+2,-2\) in the corresponding two states, for \(n\ge\max(6,\lfloor L_w/8\rfloor)=40\). The source is \(QH_{\rm ph}(t)Pu_{\rm ph}(t)=C^\dagger Pu_{\rm on}(t)\), not the radial reference. Both states use \(L_w=321\), \(z_1=1.1\), \(\Omega/J_*=1.21\), \(\varphi_1=0.49\pi\), \(\varphi_2=0\), direct \(20~{\rm G}\) effective couplings, and onsite cutoff \(M=16\); \(z_2=0,0.1\) distinguishes the waveforms. The chain is closed, with its physical left boundary and no absorber.}}
\label{fig:sm_source_response}
\end{figure}

{\color{black}Two distinct envelopes are relevant:
\begin{align}
\mathcal Q_n^{\rm snap}(t_0)
&=\left|\sum_m q_{n,m}e^{-im\Omega t_0}\right|,
&
\mathcal Q_n
&=\left(\sum_m|q_{n,m}|^2\right)^{1/2},\\
\mathcal S_n^{\rm snap}(t_0)
&=\left|\sum_m f_{n,m}e^{-im\Omega t_0}\right|,
&
\mathcal S_n
&=\left(\sum_m|f_{n,m}|^2\right)^{1/2}.
\label{eq:sm_source_envelopes}
\end{align}
The right-hand quantities are period-rms envelopes, not averages of the absolute amplitudes. The guide envelopes are dimensionless; the source envelopes have angular-frequency units and are expressed in units of \(J_*\) in dimensionless calculations. Figure~\ref{fig:sm_source_response} compares the rms source and response, each divided by its own value at \(n_0=\max(12,\lfloor L_w/6\rfloor)=53\) for \(L_w=321\). Main-text Fig.~2(c) instead compares snapshot profiles normalized at \(n_0=108\). The two figures use the same Floquet modes, with the envelopes and normalization sites specified above. Normalizing the source and response separately allows their shapes to be compared, while discarding their relative magnitudes.}

{\color{black}The geometric curve is only a dimensionless reference:
\begin{equation}
K_{\rm geo}(n)
=\frac{R_n^{-3}}{R_{n_0}^{-3}},\qquad
R_n^2=\frac{s^2}{2}+[h+(n-1)a_w]^2.
\label{eq:sm_source_proxy}
\end{equation}
Here \(s,h,a_w\) are the lengths specified in Sec.~VII. Before normalization, \(R_n^{-3}\) has units of inverse cubic length. The source \(f_{n,m}\) also depends on the Floquet state. A matching geometric slope therefore does not identify how much of the tail comes from the source: long-range guide propagation can produce an algebraic response even for a compact source.}

\subsection{\texorpdfstring{{\color{black}Long-range contribution of the guide resolvent}}{Long-range contribution of the guide resolvent}}

{\color{black}For the homogeneous inverse-cube reference guide on the full line,}
\[
E_w(k)=\omega_w+2J_w\sum_{d\ge1}\frac{\cos(dk)}{d^3},
\qquad C_3(k)\equiv\sum_{d\ge1}\frac{\cos(dk)}{d^3},
\qquad e_0=\omega_w+2J_w\zeta(3).
\]
{\color{black}Here \(k\) is dimensionless, \(\omega_w=0.70J_*\), and \(J_w\) is the signed hopping. The identity \(C_3''(k)=\log[2|\sin(k/2)|]\) gives}
\[
E_w(k)=e_0+J_w k^2\log|k|-\frac32J_wk^2+O(k^4).
\]
{\color{black}For \(E\notin\sigma(H_w^{\rm line})\), define}
\[
G_n^{\rm line}(E)=\frac{1}{2\pi}\int_{-\pi}^{\pi}
\frac{e^{ikn}}{E-E_w(k)}\,dk.
\]
{\color{black}The nonanalytic term in the resolvent is \(J_w k^2\log|k|/(E-e_0)^2\); its Fourier coefficient yields}
\begin{equation}
G_n^{\rm line}(E)\sim\frac{J_w}{(E-e_0)^2}|n|^{-3}.
\label{eq:sm_bulk_resolvent_tail}
\end{equation}
{\color{black}This full-line result shows how long-range propagation produces an algebraic response. The response coefficient in the half-line system also depends on the boundary, source, and pair-derived hopping. The plotted profiles are obtained from the finite guide matrix built from those hoppings.}

\subsection{\texorpdfstring{{\color{black}Source and propagation contributions to the semi-infinite tail}}{Source and propagation contributions to the semi-infinite tail}}

{\color{black}For the half-line \(n\ge1\), assume a real symmetric homogeneous guide with \(t_d=J_\infty d^{-3}+o(d^{-3})\), and a nonoscillatory off-band response \(f_{n,m}=A_m n^{-3}+o(n^{-3})\), \(q_{n,m}=B_m n^{-3}+o(n^{-3})\) that is absolutely summable. Choose \(h(n)\to\infty\) with \(h(n)/n\to0\) and split the hopping sum into \(1\le r\le h\), \(0<|r-n|\le h\), and the disjoint remainder. After multiplication by \(n^3\), the first part tends to \(J_\infty\mathcal M_m\), while the second tends to \(2B_m\sum_{d\ge1}t_d\). The inverse-cube bounds make the remainder vanish. Under these assumptions, substitution into \((E_m-H_w)q_m=f_m\) gives}
\begin{equation}
\left(E_m-\omega_w-2\sum_{d\ge1}t_d\right)B_m
 =A_m+J_\infty\mathcal M_m,
\qquad \mathcal M_m=\sum_{n\ge1}q_{n,m}.
\label{eq:sm_general_tail_balance}
\end{equation}
{\color{black}The direct source coefficient \(A_m\) and the propagation contribution \(J_\infty\mathcal M_m\) must be added as complex amplitudes, including their relative phase. This relation assumes the asymptotic forms stated above. The leading inverse-cube term vanishes if the two contributions on the right cancel.}

\subsection{\texorpdfstring{{\color{black}Tail coefficients of the Floquet state}}{Tail coefficients of the Floquet state}}

{\color{black}The complex coefficients are extracted independently by fitting}
\[
n^3 f_{n,m}=A_m+a_m^{(1)}n^{-1}+a_m^{(2)}n^{-2},\qquad
n^3 q_{n,m}=B_m+b_m^{(1)}n^{-1}+b_m^{(2)}n^{-2}.
\]
{\color{black}The fits use the interior interval \(\max(10,\lfloor L_w/6\rfloor)\le n\le\lfloor2L_w/3\rfloor\); for \(L_w=641\), this is \(106\le n\le427\). Define \(\mathcal M_m^{(L)}=\sum_{n=1}^{L}q_{n,m}\). The finite-guide moment is evaluated directly from the solved state rather than obtained by rearranging Eq.~(\ref{eq:sm_general_tail_balance}); the ratios \(J_\infty\mathcal M_m^{(L)}/A_m\) are approximately \(+0.0896\) for \(m=2\) and \(-0.0484\) for \(m=-2\) in the single-tone state. The opposite signs reflect constructive and destructive interference in the bound response. As with the state itself, the tail coefficients are combined coherently over the retained harmonics. Extending this identity to an infinite asymptotic expansion requires a uniform summable harmonic bound; the values below are finite-distance estimates from the converged truncated states.}

For the same single-tone states, the relative coefficient mismatch \(\left|D_mB_m-A_m-J_\infty\mathcal M_m^{(L_w)}\right|/|D_mB_m|\), where \(D_m=E_m-\omega_w-2\sum_{d\ge1}t_d\), decreases from \(1.38\times10^{-4}\) to \(3.49\times10^{-5}\) for \(m=2\), and from \(1.89\times10^{-4}\) to \(5.65\times10^{-5}\) for \(m=-2\), as \(L_w\) increases from 321 to 641. Here the coefficients are fitted independently, and the finite-guide moment is obtained by summing the solved state.

The snapshot and rms tail coefficients are
\begin{equation}
 B_{\rm snap}(t_0)=\sum_m B_m e^{-im\Omega t_0},\qquad
 B_{\rm rms}=\left(\sum_m|B_m|^2\right)^{1/2}.
\label{eq:sm_snapshot_tail_coefficient}
\end{equation}
{\color{black}At \(L_w=641\), the fitted magnitudes are \(|B_{\rm snap}(0)|\simeq0.2766\), \(B_{\rm rms}\simeq0.2574\) for the single tone and \(0.2874,0.2623\) for the biharmonic waveform. Their difference comes from interference between Fourier harmonics during a drive period. The size and fitting-window dependence in Sec.~VIII is consistent with the combined source and propagation contributions. These are estimates from finite distances; their exact infinite-guide limits remain undetermined.}

\section{Atomic input and ARC pair-state extraction}
\label{sec:sm_arc_input}

{\color{black}The static atomic input uses \({}^{87}{\rm Rb}\) and the ordered target pair states
\[
\ket{DP}=\ket{61D_{3/2},m_j=3/2;\,62P_{1/2},m_j=1/2},
\qquad \ket{PD},
\]
with the atoms exchanged in \(\ket{PD}\). The quantization axis is \(+z\), and the field \(B=20~{\rm G}\) is passed to ARC as \(B_z=0.002~{\rm T}\). ARC 3.10.2 uses \(n{\rm Range}=1\), \(l{\rm range}=3\), and \(\Delta E=2~{\rm GHz}\) for the basis used to compute the couplings; the larger reference basis uses \(n{\rm Range}=2\), \(l{\rm range}=4\), and \(\Delta E=5~{\rm GHz}\).}

\Needspace{6\baselineskip}
\subsection{\texorpdfstring{{\color{black}Fixed-energy pair reduction}}{Fixed-energy pair reduction}}

{\color{black}Let \(P_{\rm at}\) project onto the two bare pair states \(DP,PD\), and let \(Q_{\rm at}=I-P_{\rm at}\) within the finite atomic basis. For the block matrix}
\begin{equation}
\mathcal A=
\begin{pmatrix}\mathcal A_{PP}&\mathcal A_{PQ}\\
\mathcal A_{QP}&\mathcal A_{QQ}\end{pmatrix},
\end{equation}
{\color{black}eliminating \(q=Q_{\rm at}v\) from \(\mathcal A v=Ev\) gives the exact energy-dependent equation}
\begin{equation}
\mathcal A_{\rm eff}(E)p=Ep,\qquad
\mathcal A_{\rm eff}(E)=\mathcal A_{PP}
 +\mathcal A_{PQ}(E-\mathcal A_{QQ})^{-1}\mathcal A_{QP}.
\label{eq:sm_atomic_schur}
\end{equation}
{\color{black}The effective spin model instead evaluates the signed exchange at the prescribed bare-block midpoint}
\begin{equation}
 E_0=\tfrac12\operatorname{Re}\operatorname{Tr}\mathcal A_{PP},\qquad
 J_{ab}^{\rm cyc}=\left[\mathcal A_{\rm eff}(E_0)\right]_{DP,PD},
\label{eq:sm_atomic_midpoint}
\end{equation}
{\color{black}with the conjugate entry on the opposite triangle. This fixed-energy approximation defines the exchange in the effective spin Hamiltonian; its static accuracy is characterized below.}

{\color{black}The infinite contact sequence is defined with one fixed magnetic-sublevel basis, including contacts whose polar angle approaches or rounds to zero. The same 260 states are retained throughout the contact sequence. Exactly axial guide--guide pairs retain their separate 34-state basis. The radius and polar angle are rounded to \(10^{-6}\,\mu{\rm m}\) and \(10^{-8}\) radians, respectively. The retained angular matrix-element cutoff is \(10^{-5}\), and the ARC calculation retains the \(10^{-8}\) degeneracy offset in the source code. These conventions are kept fixed in the finite matrices and their infinite continuation.}

{\color{black}The accuracy of the pair reduction is quantified using normalized full pair eigenvectors and the same midpoint matrix. For a selected pair, the largest individual outside-target weight is \(1-\min(w_1,w_2)\), with \(w_a=\|P_{\rm at}v_a\|^2\), and the relative splitting discrepancy is}
\begin{equation}
\epsilon_{\rm split}=
\frac{\bigl|\,|\lambda_2-\lambda_1|-|E_2-E_1|\,\bigr|}{|E_2-E_1|},
\label{eq:sm_atomic_splitting_error}
\end{equation}
{\color{black}where \(\lambda_a\) are the midpoint-matrix eigenvalues and \(E_a\) the corresponding full pair eigenvalues.}

\begin{center}
\begin{tabular}{lrrr}
\toprule
Pair & Reference dimension & Outside-target weight & Splitting discrepancy\\
\midrule
First contact & 1188 & \(8.49\%\) & \(7.27\%\)\\
Plaquette edge & 1188 & \(0.602\%\) & \(0.568\%\)\\
Nearest guide pair & 134 & \(0.00306\%\) & \(0.00302\%\)\\
\bottomrule
\end{tabular}
\end{center}

{\color{black}These quantities measure the static pair approximation. The many-site dynamics is governed by the resulting effective spin Hamiltonian. All 300 finite-device pairs and the longer axial contacts are evaluated directly at their actual distances and angles, with the zero-regulator Hermitian limit used for the Floquet matrices. The ARC calculation supplies static exchange parameters for the effective spin Hamiltonian; a driven multilevel atomic realization would require additional drive matrix elements that are not included in this effective model.}

\section{Axial geometry and pair-exchange matrix}
\label{sec:sm_selected_geometry}

{\color{black}The waveguide passes through the plaquette center. The dimensions quoted in the main text are rounded for presentation. The numerical operator and the bounds in Sec.~IV use the original coordinate parameters, including \(s=12.412808860824748~\mu{\rm m}\), \(h=3.948921113913186~\mu{\rm m}\), and \(a_w=12.778131633247837~\mu{\rm m}\). The coordinates, ordered consistently with the single-excitation basis, are
\begin{align}
\bm r_{p_0}&=(-s/2,-s/2,0),&
\bm r_{p_1}&=(s/2,-s/2,0),\\
\bm r_{p_2}&=(s/2,s/2,0),&
\bm r_{p_3}&=(-s/2,s/2,0),\\
\bm r_{s_n}&=(\delta a_w,0,h+(n-1)a_w),&& n=1,\ldots,L_w .
\label{eq:sm_coordinates}
\end{align}
The axial calculation has \(\delta=0\); transverse displacement is discussed in Sec.~IX. All six plaquette links, \(4L_w\) contact links, and \(L_w(L_w-1)/2\) guide links are retained. Thus \(L_w=21\) gives 25 sites and 300 distinct pairs. The coordinate values above and the rounding conventions in Sec.~VI specify the numerical input. These digits fix the numerical geometry; they do not specify experimental precision or the accuracy of the atomic reduction.}

{\color{black}For a pair displacement \(\bm R_{ab}=\bm r_b-\bm r_a\), the pair reduction uses its length \(R_{ab}\), polar angle relative to \(+z\), and azimuth. The exchange retains its sign and phase in the upper triangle, with its Hermitian conjugate in the lower triangle. Equivalent axial pairs share the same calculated exchange; different pair distances are evaluated separately, rather than generated by rescaling a single finite-distance Schur matrix. Along the guide the exchange depends only on the separation \(d=|n-r|\). Along the contacts both distance and polar angle change with \(n\).}

{\color{black}The static-channel projection in Eq.~(\ref{eq:sm_static_channel}) is a four-site amplitude sum. For identical axial contacts \(C_{jn}=c_n^{(0)}\),
\begin{equation}
J_{\ell_s n}^{(0)}
=\frac{c_n^{(0)}}{2}\sum_{j=0}^3e^{-i\ell_s\pi j/2}
=2c_n^{(0)}\delta_{\ell_s,0}^{(4)}.
\label{eq:sm_axial_static_channel}
\end{equation}
The other angular components vanish by exact axial coupling symmetry. The direct pair matrix supplies an approximate microscopic value of the nonzero contact. The finite atomic bases and the fixed-energy reduction are specified in Sec.~VI. Main-text Fig.~3 displays the geometry and angular-channel response; Fig.~\ref{fig:sm_sambe_channels} addresses the distinct angular-harmonic content of the Floquet eigenstate.}

\section{Branch tracking and algebraic tail fits}
\label{sec:sm_branch_tail_fits}

The same branch is followed across guide lengths by maximum overlap with the state at the preceding length, avoiding switches between nearby folded quasienergies. At each step, the shorter vector is embedded in the longer Hilbert space before the overlaps are evaluated. Main-text Fig.~2(a,b) shows the spectral distances and local weights along this branch.

\subsection{\texorpdfstring{{\color{black}Harmonic and propagation convergence}}{Harmonic and propagation convergence}}

{\color{black}The harmonic cutoff and guide length are varied independently, using
\(M=8,12,16\) and \(L_w=21,41,81,161,321,641\), with a 1024-point periodic reconstruction.}

{\color{black}Increasing \(M\) from 12 to 16 changes \(P_{\chi_-}\) by less than \(2\times10^{-12}\), while the periodic norm discrepancy at \(M=16\) is below \(6\times10^{-15}\).}

{\color{black}For the folded-spectrum spacing we use the 1024-slice one-period propagator; at \(L_w=321\) its tracked quasienergy differs from the \(M=16\) Sambe value by \(4.9\times10^{-7}J_*\) for the single tone and \(6.2\times10^{-7}J_*\) for the biharmonic drive. At this length,
\begin{center}
\begin{tabular}{lrrrrr}
\toprule
Waveform & \(\epsilon/J_*\) & \(P_{\rm plaq}\) & \(P_{\chi_-}(0)\) & IPR & \(\Delta_{\rm ext}/J_*\)\\
\midrule
Single tone & \(0.09057\) & \(0.8726\) & \(0.6151\) & \(0.2698\) & \(5.85\times10^{-4}\)\\
Biharmonic & \(0.1215\) & \(0.8705\) & \(0.6211\) & \(0.2655\) & \(2.26\times10^{-3}\)\\
\bottomrule
\end{tabular}
\end{center}
The first four numerical columns use the same normalized Sambe mode and its phase-frame \(t=0\) reconstruction; the last uses the full propagator spectrum. At \(L_w=641\) the local quantities are unchanged at the displayed precision, while the corresponding spacings are about \(1.09\times10^{-4}\) and \(6.41\times10^{-4}\). The decreasing finite-guide spacings show the approach of folded extended levels, while Sec.~IV establishes the corresponding infinite-guide sector structure.}

\subsection{\texorpdfstring{{\color{black}Fixed waveform family}}{Fixed waveform family}}

\begin{figure}[t]
\includegraphics[width=0.90\textwidth]{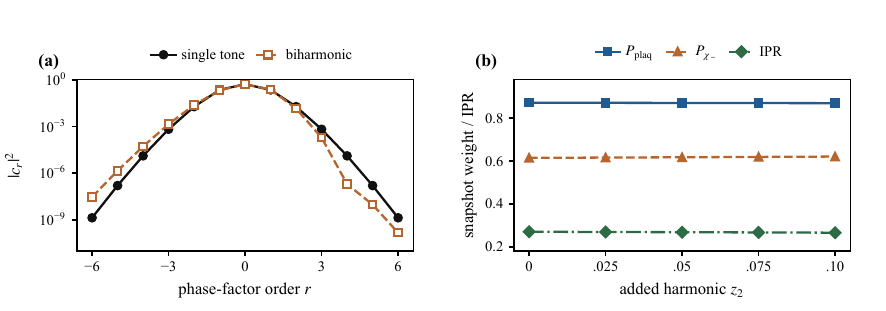}
\caption{{\color{black}Phase-factor coefficients and response within a fixed symmetry-preserving waveform family. (a) The input weights \(|c_r|^2\) follow from the phase
\(z_1\sin(x+\varphi_1)+(z_2/2)\sin(2x+\varphi_2)\), with \(x=\Omega t\), and are not Floquet-state populations. (b) The state response consists of \(P_{\rm plaq}(0)\), \(P_{\chi_-}(0)\), and IPR along \(z_2=0,0.025,0.05,0.075,0.1\) at \(L_w=321\). The direct \(20~{\rm G}\) effective couplings, geometry, onsite terms, \(\Omega/J_*=1.21\), \(z_1=1.1\), \(\varphi_1=0.49\pi\), and \(\varphi_2=0\) are fixed. The phase coefficients use the full periodic phase factor, and the normalized states use onsite cutoff \(M=16\) with coherent phase-frame reconstruction.}}
\label{fig:sm_waveform}
\end{figure}

{\color{black}The phase-factor coefficients in Eq.~(\ref{eq:sm_phase_coefficients}) and the five-point branch continuation are shown together in Fig.~\ref{fig:sm_waveform}. The former describe the imposed drive, whereas the latter show snapshot populations and IPR of its normalized Floquet eigenstate. The doubled winding of the added harmonic preserves the joint symmetry throughout this family. The intermediate points change only \(z_2\), and each state is connected by overlap with the preceding one. The microscopic effective matrix and all remaining drive parameters are held fixed.}

\subsection{\texorpdfstring{{\color{black}Interior windows and finite-distance exponents}}{Interior windows and finite-distance exponents}}

{\color{black}The guide index starts at \(n=1\), excluding the four plaquette sites. The snapshot exponent is fitted from
\begin{equation}
\log\mathcal Q_n^{\rm snap}(0)=\log A_{\rm fit}-\alpha\log n .
\label{eq:sm_amplitude_fit}
\end{equation}
The amplitude \(A_{\rm fit}\) here is nonnegative and is distinct from the complex harmonic source coefficient \(A_m\). For a finite chain, a tail window must stay far from both the contact region and the terminal boundary. Three interior windows are used:
\begin{align}
\mathcal W_A&=[\max(6,\lfloor L_w/8\rfloor),\ \max(12,\lfloor L_w/2\rfloor)],
\label{eq:sm_fit_window_a}\\
\mathcal W_B&=[\max(6,\lfloor L_w/6\rfloor),\ \max(12,\lfloor2L_w/3\rfloor)],
\label{eq:sm_fit_window_b}\\
\mathcal W_C&=[\max(6,\lfloor L_w/4\rfloor),\ \max(12,\lfloor3L_w/4\rfloor)].
\label{eq:sm_fit_windows}
\end{align}
Fits with fewer than six sites are omitted. For comparison, the endpoint-reaching window is
\([\max(6,\lfloor L_w/3\rfloor+1),L_w]\).
At \(L_w=321\) it is \(108\le n\le321\), spanning less than half a decade and including the terminal boundary. It is not an interior infinite-chain limit.}

\begin{figure}[t]
\includegraphics[width=0.90\textwidth]{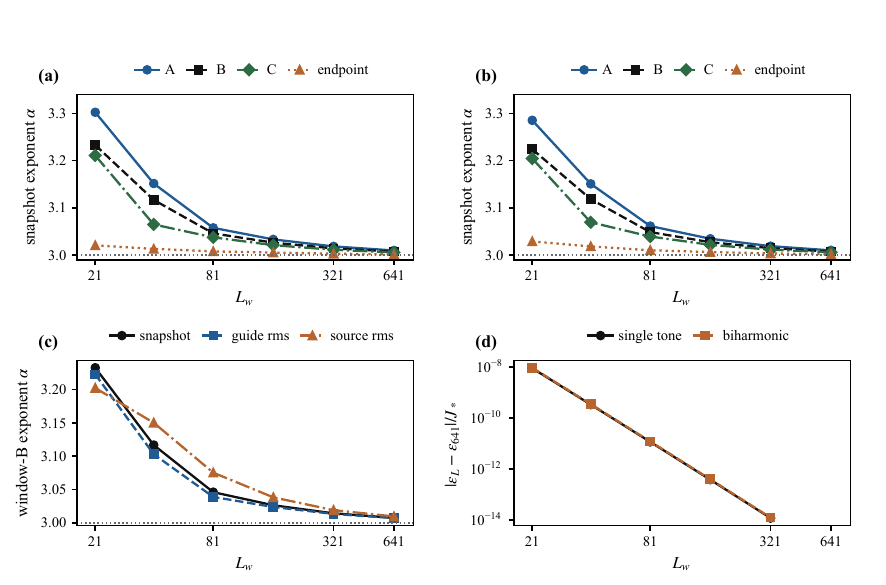}
\caption{{\color{black}Finite-size and fitting-window dependence of the algebraic envelope. (a,b) Snapshot exponents for the single-tone and biharmonic modes in the three interior windows \(\mathcal W_A,\mathcal W_B,\mathcal W_C\) of Eqs.~(\ref{eq:sm_fit_window_a})--(\ref{eq:sm_fit_windows}); an endpoint-reaching window is retained as a boundary-sensitive comparison. (c) Single-tone exponents of the snapshot guide amplitude, period-rms guide, and period-rms source, each fitted in \(\mathcal W_B\). (d) Finite-size quasienergy difference \(|\epsilon_{L_w}-\epsilon_{641}|/J_*\) for both waveforms. The identically zero \(L_w=641\) point is omitted on the logarithmic axis; the reference is a finite chain, not an asymptotic extrapolation. Both \(z_2=0\) and \(0.1\) use direct \(20~{\rm G}\) effective couplings with unchanged geometry and drive parameters, \(\Omega/J_*=1.21\), \(z_1=1.1\), \(\varphi_1=0.49\pi\), and \(\varphi_2=0\). The length sequence is \(L_w=21,41,81,161,321,641\), with onsite cutoff \(M=16\). The spread across windows shows the finite-distance sensitivity of the extracted exponent.}}
\label{fig:sm_tail_branch}
\end{figure}

{\color{black}The fitted snapshot exponents are
\begin{center}
\begin{tabular}{llrrr}
\toprule
\(L_w\) & Window & Sites \(n\) & Single-tone \(\alpha\) & Biharmonic \(\alpha\)\\
\midrule
321 & \(\mathcal W_A\) & \(40\)--\(160\) & \(3.018\) & \(3.019\)\\
321 & \(\mathcal W_B\) & \(53\)--\(214\) & \(3.014\) & \(3.015\)\\
321 & \(\mathcal W_C\) & \(80\)--\(240\) & \(3.011\) & \(3.012\)\\
641 & \(\mathcal W_A\) & \(80\)--\(320\) & \(3.010\) & \(3.010\)\\
641 & \(\mathcal W_B\) & \(106\)--\(427\) & \(3.008\) & \(3.008\)\\
641 & \(\mathcal W_C\) & \(160\)--\(480\) & \(3.006\) & \(3.006\)\\
\bottomrule
\end{tabular}
\end{center}
Figure~\ref{fig:sm_tail_branch} displays the separate windows and size dependence. Main-text exponent markers use \(\mathcal W_B\), with whiskers spanning the minimum and maximum over the three windows. The spread among the interior windows measures the sensitivity of the fitted exponent to the chosen distance interval. As the guide grows, all three interior fits approach an amplitude exponent of three. The endpoint-reaching window is shown separately because it is more sensitive to the terminal boundary.}

{\color{black}The residual profile
\(r_n=\log\mathcal Q_n^{\rm snap}(0)-[\log A_{\rm fit}-\alpha\log n]\)
distinguishes a smooth finite-distance trend from the terminal distortion. The rms guide and rms source can be fitted separately using Eq.~(\ref{eq:sm_source_envelopes}), but they must not be substituted for the snapshot observable without changing its definition. Squaring the same positive amplitudes gives
\(\log|\psi_n|^2=2\log|\psi_n|\), so an identical least-squares window yields exactly twice the fitted exponent. This factor of two follows from squaring the amplitudes and provides no additional evidence for their spatial decay. If a nonzero \(n^{-3}\) amplitude asymptote holds, its \(n^{-6}\) probability is summable in one dimension. Section~V relates the coefficients to the source and guide response under explicit asymptotic assumptions; a finite-window fit alone does not determine the asymptotic law of a general long-range system.}

\FloatBarrier
\Needspace{7\baselineskip}
\section{Transverse displacement and angular-channel mixing}
\label{sec:sm_c4_breaking}

{\color{black}A transverse shift \(\Delta x=\delta a_w\) changes the pair distances and angles in Eq.~(\ref{eq:sm_coordinates}) and breaks the axial symmetry. The static unwanted-channel ratio is
\begin{equation}
\eta_{C_4}(\delta)
=\frac{\displaystyle\sum_{n=1}^{8}
\sum_{\ell\ne0}|J_{\ell n}^{(0)}(\delta)|^2}
{\displaystyle\sum_{n=1}^{8}|J_{0n}^{(0)}(\delta)|^2}.
\label{eq:sm_displacement_metric}
\end{equation}
The ratio \(\eta_{C_4}\) measures the static angular admixture generated by the transverse displacement. Each point is evaluated directly at the displaced pair geometry at a field of \(20\) G.}

{\color{black}Expanding the forbidden angular amplitudes about the axial point gives
\[
J_{\ell n}^{(0)}(\delta)
=\delta\,\partial_\delta J_{\ell n}^{(0)}(0)+O(\delta^2)
\quad(\ell\ne0).
\]
Provided the axial denominator is nonzero, substitution into Eq.~(\ref{eq:sm_displacement_metric}) yields
\begin{equation}
\eta_{C_4}(\delta)=c_\delta\delta^2+O(\delta^4),\qquad
c_\delta=
\frac{\sum_{n=1}^8\sum_{\ell\ne0}
|\partial_\delta J_{\ell n}^{(0)}(0)|^2}
{\sum_{n=1}^8|J_{0n}^{(0)}(0)|^2}.
\label{eq:sm_displacement_quadratic}
\end{equation}
Reflection symmetry removes the cubic term, as shown below, leaving a fourth-order remainder whose coefficient depends on the pair exchange and geometry.}

{\color{black}For a displacement along \(+x\), reflection across the \(xz\) plane groups the four contacts at guide site \(n\) into
\begin{equation}
(C_{0n},C_{1n},C_{2n},C_{3n})
=(a_n(\delta),b_n(\delta),b_n(\delta),a_n(\delta)).
\label{eq:sm_displacement_contacts}
\end{equation}
Here \(a_n,b_n\) denote the two static contact amplitudes at this displacement, not the time-dependent state coefficients. Applying the four-site projection yields
\begin{equation}
J_{0n}^{(0)}=a_n+b_n,\qquad
J_{+1,n}^{(0)}=\frac{1+i}{2}(a_n-b_n),\qquad
J_{-1,n}^{(0)}=\frac{1-i}{2}(a_n-b_n),\qquad
J_{2n}^{(0)}=0.
\label{eq:sm_displacement_channels}
\end{equation}
Consequently,
\begin{equation}
\eta_{C_4}(\delta)
=\frac{\sum_{n=1}^8|a_n(\delta)-b_n(\delta)|^2}
{\sum_{n=1}^8|a_n(\delta)+b_n(\delta)|^2}.
\label{eq:sm_displacement_even_ratio}
\end{equation}
Reversing the displacement exchanges the two groups,
\(a_n(-\delta)=b_n(\delta)\) and \(b_n(-\delta)=a_n(\delta)\).
The numerator and denominator are therefore even. If the underlying pair exchange varies smoothly with displacement and the axial denominator is nonzero, the leading odd contact difference is linear in \(\delta\), while the contact sum has only even corrections. Equation~(\ref{eq:sm_displacement_quadratic}) then has an \(O(\delta^4)\) remainder. The analytic expansion applies to this smooth dependence; the plotted response is evaluated directly at the discrete pair geometry.}

{\color{black}The directly evaluated static ratios are
\begin{center}
\begin{tabular}{crrrrrrr}
\toprule
\(\delta\) & 0 & 0.01 & 0.02 & 0.05 & 0.10 & 0.15 & 0.20\\
\midrule
\(\eta_{C_4}\) & 0 & \(1.4316\times10^{-4}\) & \(5.7307\times10^{-4}\) & \(0.0036006\) & \(0.014718\) & \(0.034616\) & \(0.066232\)\\
\bottomrule
\end{tabular}
\end{center}
The axial value vanishes by symmetry; the residual numerical value at \(\delta=0\) is of order \(10^{-32}\). All entries use the same \(20~{\rm G}\) field, zero inverse regulator, fixed midpoint, and direct pair evaluation at the displaced geometry.}

\begin{figure}[t]
\includegraphics[width=0.90\textwidth]{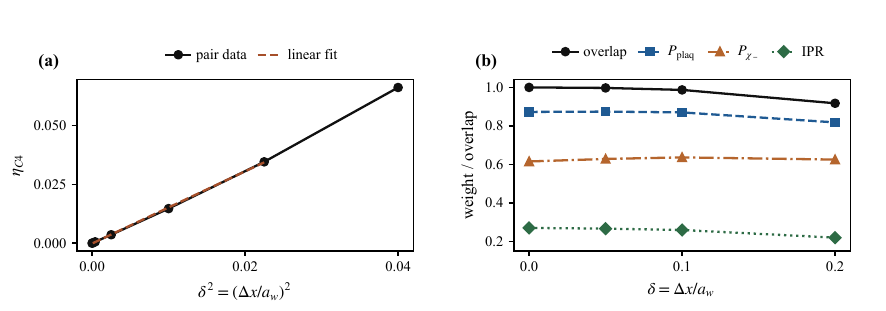}
\caption{{\color{black}Response to a transverse displacement \(\Delta x=\delta a_w\) with directly evaluated pair couplings. (a) Static unwanted-angular-channel ratio \(\eta_{C_4}\) of Eq.~(\ref{eq:sm_displacement_metric}), summed over the first eight guide sites, versus \(\delta^2\). The seven points extend through \(\delta=0.20\); the dashed line is the equal-weight, free-intercept fit over the six points with \(\delta\le0.15\), not a fit over the entire plotted window. (b) Full-state snapshot plaquette weight, target-chirality weight, IPR, and squared axial-reference overlap \(\mathcal O_{\rm ax}\) at \(\delta=0,0.05,0.10,0.20\). The finite guide has \(L_w=161\), open ends, and no terminal loss. Each displaced matrix uses direct zero-regulator pair reduction at \(B=20~{\rm G}\), with the geometry of Sec.~VII shifted along \(x\), and unchanged imposed onsites. The single-tone drive has \(z_1=1.1\), \(z_2=0\), \(\Omega/J_*=1.21\), and \(\varphi_1=0.49\pi\); all plaquette-incident bonds are driven. The states are obtained from the full-period fourth-order propagator with 1024 steps, with no sector reduction, and are evaluated in the phase frame at \(t_0=0\). The overlap in panel (b) is with the same \(\delta=0\) state, not the adjacent displacement. Panel (a) measures static angular mixing, while panel (b) follows the finite Floquet branch.}}
\label{fig:sm_c4}
\end{figure}

{\color{black}To quantify finite-window curvature, we fit the static data with equal weights to
\begin{equation}
\eta_{C_4}^{\rm fit}(\delta)=a_{\rm fit}\delta^2+b_{\rm fit},
\label{eq:sm_displacement_fit}
\end{equation}
allowing the intercept to vary. The six points in \(0\le\delta\le0.15\) give \(a_{\rm fit}=1.53456\), \(b_{\rm fit}=-1.37637\times10^{-4}\), and \(R^2=0.999635\). This is the fit drawn in Fig.~\ref{fig:sm_c4}(a). Including the seventh point at \(\delta=0.20\) changes the fit to \(a_{\rm fit}=1.63874\), \(b_{\rm fit}=-5.48947\times10^{-4}\), and \(R^2=0.998276\). The change in slope after the \(\delta=0.20\) point is included is consistent with the \(O(\delta^4)\) curvature of Eq.~(\ref{eq:sm_displacement_quadratic}); the exact axial value remains \(\eta_{C_4}(0)=0\).}

{\color{black}At nonzero displacement the Floquet state is obtained from the full closed Hamiltonian, since the exact \(\kappa=2\) block is no longer invariant. We use \(L_w=161\) and \(\delta=0,0.05,0.10,0.20\), keeping the single-tone waveform \(z_1=1.1\), \(z_2=0\), \(\Omega/J_*=1.21\), and \(\varphi_1=0.49\pi\), as well as all imposed onsite terms and geometry parameters, fixed. Every plaquette-incident bond, including the internal plaquette bonds, is transformed by \(F(t)H_0F^\dagger(t)\). No symmetry-sector truncation or quarter-period reduction is used.}

{\color{black}Doubling the 1024-step full-period integration and comparing with an independent full-period integrator change the observables by less than \(4.5\times10^{-9}\) and give a one-period closure change below \(1.4\times10^{-10}\).}

{\color{black}The reference at \(\delta=0\) is the axial branch identified and tracked in Sec.~VIII. At each nonzero displacement we follow the state with maximum squared overlap with this fixed axial reference. Tracking instead by overlap with the immediately preceding displacement selects the same state throughout the plotted interval. The displayed overlap is then
\begin{equation}
\mathcal O_{\rm ax}(\delta)
=|\braket{u_\delta(0)}{u_0(0)}|^2.
\label{eq:sm_axial_overlap}
\end{equation}
All vectors have full-system unit normalization and the same phase-frame time origin \(t_0=0\). In particular, at \(\delta=0.20\) the adjacent overlap is \(0.9602\), whereas the axial overlap is \(0.9176\).}

{\color{black}The full-state snapshot observables are
\begin{center}
\begin{tabular}{c@{\quad}r@{\quad}r@{\quad}r@{\quad}r@{\quad}r}
\toprule
\(\delta\) & \(\epsilon/J_*\) & \(P_{\rm plaq}(0)\) & \(P_{\chi_-}(0)\) & IPR & \(\mathcal O_{\rm ax}\)\\
\midrule
0 & \(0.09057325\) & \(0.87264801\) & \(0.61513107\) & \(0.26982286\) & \(1\)\\
0.05 & \(0.09032441\) & \(0.87434684\) & \(0.62828397\) & \(0.26611163\) & \(0.99732476\)\\
0.10 & \(0.08995049\) & \(0.87014713\) & \(0.63636266\) & \(0.25899639\) & \(0.98709730\)\\
0.20 & \(0.09429357\) & \(0.81827736\) & \(0.62502404\) & \(0.21928812\) & \(0.91757770\)\\
\bottomrule
\end{tabular}
\end{center}
Figure~S5 therefore follows the finite Floquet branch as displacement breaks its protecting symmetry. The infinite-guide embedded eigenstate established in Sec.~IV applies at the axial point \(\delta=0\).}

\end{document}